\documentclass[11pt, a4paper]{article}
\usepackage{jcappub}
\usepackage{orcidlink}
\usepackage{multirow}
\usepackage{wrapfig}
\usepackage{longtable}
\usepackage{fp}
\usepackage{natbib}
\usepackage{color}
\usepackage{amsmath,amsfonts,bm}
\usepackage{hyperref}
\usepackage[T1]{fontenc}
\usepackage{graphicx}	
\usepackage{cleveref}
\usepackage[mathlines]{lineno}

\graphicspath{{figs/}}

\defcitealias{zahn12}{Z12}
\defcitealias{george15}{G15}
\defcitealias{reichardt21}{R21}
\defcitealias{chaubal25}{C25}

\newcommand{\farcm}{\ensuremath{^\prime}}

\newcommand{\be}{\begin{equation}}
\newcommand{\ee}{\end{equation}}
\newcommand{\bea}{\begin{eqnarray}}
\newcommand{\eea}{\end{eqnarray}}

\newcommand{\lcdm}{$\Lambda$CDM}

\newcommand{\planck}{{\it Planck}}

\newcommand{\sqdeg}{\mbox{deg$^2$}}
\newcommand{\chisq}{\ensuremath{\chi^2}}
\newcommand{\delchisq}{\ensuremath{\Delta\chi^2}}

\newcommand{\ltsima}{$\; \buildrel < \over \sim \;$}
\newcommand{\ltsim}{\lower.5ex\hbox{\ltsima}}

\newcommand{\comment}[1]{}

\newcommand{\uksq}{\ensuremath{\mu {\rm K}^2}}
\newcommand{\dksz}{\ensuremath{D_{3000}^{\rm kSZ}}}
\newcommand{\dpksz}{\ensuremath{D_{3000}^{\rm p-kSZ}}}
\newcommand{\dhksz}{\ensuremath{D_{3000}^{\rm h-kSZ}}}
\newcommand{\dtsz}{\ensuremath{D_{3000}^{\rm tSZ}}}

\newcommand{\dtszagora}{\ensuremath{4.28\pm0.37\, \mu{\rm K}^2}}

\newcommand{\dkszagora}{\ensuremath{3.96\pm0.82\, \mu{\rm K}^2}}

\newcommand{\delzwhkszpriordzagora}{\ensuremath{3.9^{+3.5}_{-2.1}}}

\newcommand{\limdelzwhkszpriordzagora}{\ensuremath{<\,11}}

\newcommand{\delzwhkszpriordzamberagora}{\ensuremath{7.0^{+4.6}_{-3.3}}}

\newcommand{\drgonefiftyagora}{\ensuremath{1.44\pm0.18\, \mu{\rm K}^2}}

\newcommand{\drgninetyfiveagora}{\ensuremath{7.22\pm0.29\, \mu{\rm K}^2}}

\newcommand{\dcibponefiftyagora}{\ensuremath{3.56\pm0.65\, \mu{\rm K}^2}}
\newcommand{\dcibptwotwentyagora}{\ensuremath{25.7\pm6.0\, \mu{\rm K}^2}}

\newcommand{\dcibclaonefiftyagora}{\ensuremath{5.16\pm0.65\, \mu{\rm K}^2}}
\newcommand{\dcibclatwotwentyagora}{\ensuremath{64.1\pm6.2\, \mu{\rm K}^2}}

\newcommand{\dcibclbonefiftyagora}{\ensuremath{1.40\pm0.11\, \mu{\rm K}^2}}
\newcommand{\dcibclbtwotwentyagora}{\ensuremath{17.4\pm1.3\, \mu{\rm K}^2}}

\newcommand{\dcibtotonefiftyagora}{\ensuremath{10.11\pm0.22\, \mu{\rm K}^2}}
\newcommand{\dcibtottwotwentyagora}{\ensuremath{107.2\pm1.7\, \mu{\rm K}^2}}

\newcommand{\tszcibagora}{\ensuremath{0.091\pm0.025}}
\newcommand{\dtszbase}{\ensuremath{5.47\pm0.20\, \mu{\rm K}^2}}

\newcommand{\limkszbase}{\ensuremath{<\,1.4\, \mu{\rm K}^2}}

\newcommand{\dtszagorawbasetsz}{\ensuremath{4.79\pm0.38\, \mu{\rm K}^2}}

\newcommand{\dkszagorawbasetsz}{\ensuremath{2.42\pm0.89\, \mu{\rm K}^2}}

\newcommand{\dtszagorawbasetszcib}{\ensuremath{5.13\pm0.24\, \mu{\rm K}^2}}

\newcommand{\dkszagorawbasetszcib}{\ensuremath{1.89\pm0.51\, \mu{\rm K}^2}}

\newcommand{\dtszpowerlawphys}{\ensuremath{4.98\pm0.35\, \mu{\rm K}^2}}

\newcommand{\dkszpowerlawphys}{\ensuremath{1.98\pm0.87\, \mu{\rm K}^2}}

\newcommand{\delzwhkszpriordzpowerlawphys}{\ensuremath{0.8^{+1.5}_{-0.6}}}

\newcommand{\limdelzwhkszpriordzpowerlawphys}{\ensuremath{<\,4.2}}

\newcommand{\delzwhkszpriordzamberpowerlawphys}{\ensuremath{1.8^{+2.5}_{-1.3}}}

\newcommand{\limdelzwhkszpriordzamberpowerlawphys}{\ensuremath{<\,7.0}}

\newcommand{\dkszfourkpowerlawphys}{\ensuremath{1.66\pm0.84\, \mu{\rm K}^2}}
\newcommand{\dkszfivekpowerlawphys}{\ensuremath{1.45\pm0.87\, \mu{\rm K}^2}}
\newcommand{\dkszsixkpowerlawphys}{\ensuremath{1.28\pm0.93\, \mu{\rm K}^2}}
\newcommand{\dksztwokpowerlawphys}{\ensuremath{2.4\pm1.5\, \mu{\rm K}^2}}
\newcommand{\dtsztwokpowerlawphys}{\ensuremath{4.89\pm0.47\, \mu{\rm K}^2}}
\newcommand{\dtszfourkpowerlawphys}{\ensuremath{4.71\pm0.40\, \mu{\rm K}^2}}
\newcommand{\dtszfivekpowerlawphys}{\ensuremath{4.34\pm0.46\, \mu{\rm K}^2}}
\newcommand{\dtszsixkpowerlawphys}{\ensuremath{3.94\pm0.51\, \mu{\rm K}^2}}

\newcommand{\drgonefiftypowerlawphys}{\ensuremath{1.25\pm0.17\, \mu{\rm K}^2}}

\newcommand{\drgninetyfivepowerlawphys}{\ensuremath{6.61\pm0.31\, \mu{\rm K}^2}}

\newcommand{\dcibponefiftypowerlawphys}{\ensuremath{6.43\pm0.59\, \mu{\rm K}^2}}
\newcommand{\dcibptwotwentypowerlawphys}{\ensuremath{54.1\pm5.7\, \mu{\rm K}^2}}

\newcommand{\tszcibpowerlawphys}{\ensuremath{0.036\pm0.021}}
\newcommand{\dcibclonefiftypowerlawphys}{\ensuremath{4.18\pm0.51\, \mu{\rm K}^2}}
\newcommand{\dcibcltwotwentypowerlawphys}{\ensuremath{54.9\pm5.3\, \mu{\rm K}^2}}

\newcommand{\dcibtotninetyfivepowerlawphys}{\ensuremath{1.352\pm0.081\, \mu{\rm K}^2}}
\newcommand{\dcibtotonefiftypowerlawphys}{\ensuremath{10.61\pm0.29\, \mu{\rm K}^2}}
\newcommand{\dcibtottwotwentypowerlawphys}{\ensuremath{109.1\pm2.1\, \mu{\rm K}^2}}
\newcommand{\dtszfreecibphysagorasz}{\ensuremath{4.61\pm0.38\, \mu{\rm K}^2}}

\newcommand{\dkszfreecibphysagorasz}{\ensuremath{2.89\pm0.87\, \mu{\rm K}^2}}

\newcommand{\dtszfreecibphysbasesz}{\ensuremath{4.96\pm0.28\, \mu{\rm K}^2}}

\newcommand{\dkszfreecibphysbasesz}{\ensuremath{1.74\pm0.64\, \mu{\rm K}^2}}

\newcommand{\dtszmorecszphys}{\ensuremath{4.33\pm0.53\, \mu{\rm K}^2}}

\newcommand{\dkszmorecszphys}{\ensuremath{2.5\pm1.1\, \mu{\rm K}^2}}

\newcommand{\dksztwokmorecszphys}{\ensuremath{2.4\pm1.7\, \mu{\rm K}^2}}
\newcommand{\dkszfourkmorecszphys}{\ensuremath{2.85\pm0.69\, \mu{\rm K}^2}}
\newcommand{\dkszfivekmorecszphys}{\ensuremath{3.45\pm0.90\, \mu{\rm K}^2}}
\newcommand{\dkszsixkmorecszphys}{\ensuremath{2.3\pm1.0\, \mu{\rm K}^2}}
\newcommand{\dtsztwokmorecszphys}{\ensuremath{3.9\pm1.2\, \mu{\rm K}^2}}
\newcommand{\dtszfourkmorecszphys}{\ensuremath{3.87\pm0.36\, \mu{\rm K}^2}}
\newcommand{\dtszfivekmorecszphys}{\ensuremath{3.17\pm0.53\, \mu{\rm K}^2}}
\newcommand{\dtszsixkmorecszphys}{\ensuremath{3.07\pm0.61\, \mu{\rm K}^2}}
\newcommand{\drgonefiftymorecszphys}{\ensuremath{1.37\pm0.20\, \mu{\rm K}^2}}

\newcommand{\drgninetyfivemorecszphys}{\ensuremath{6.96\pm0.34\, \mu{\rm K}^2}}

\newcommand{\tszcibmorecszphys}{\ensuremath{0.059\pm0.033}}

\newcommand{\dcibtotonefiftymorecszphys}{\ensuremath{10.16\pm0.30\, \mu{\rm K}^2}}
\newcommand{\dcibtottwotwentymorecszphys}{\ensuremath{107.4\pm2.0\, \mu{\rm K}^2}}

\newcommand{\tittext}{SPT-3G D1: A Measurement of Secondary Cosmic Microwave Background Anisotropy Power}

\newcommand{\abstracttext}{
We report new measurements of millimeter-wave temperature power spectra in the angular multipole range $1700 \le \ell \le 11,000$ (wavelengths $13^\prime \gtrsim \lambda \gtrsim 2^\prime$).  
We use two years of data in three observing bands centered near 95, 150, and 220\,GHz from the SPT-3G receiver on the South Pole Telescope 
that cover a 1646\,\sqdeg{} region of the Southern sky. 
Using the measured power spectra, we present constraints on the thermal and kinematic Sunyaev-Zel'dovich (SZ) effects, radio galaxies, and cosmic infrared background (CIB). 
We find that inferred SZ powers are dependent on the detailed modeling of the thermal SZ-CIB correlation, and to a lesser extent on the assumed angular dependence of the SZ spectra. 
We report constraints for simulation-based model templates as well as fits where the angular dependencies of the SZ  and CIB power spectra are allowed to vary. 
In the latter case at $\ell=3000$, we find thermal SZ power at 143 GHz of $\dtsz = \dtszpowerlawphys$ and kinematic SZ power of $\dksz =\dkszpowerlawphys$. 
We use the measured kinematic SZ power to estimate the duration of reionization, noting that the reionization inferences are sensitive to the model choices and  assumed level of  homogeneous kinematic SZ power from the late-time universe. 
We find a 95\% limit on the duration from an ionization fraction of 25\% to 75\% of $\Delta^{50} z_{\rm re}  \limdelzwhkszpriordzpowerlawphys$ based on a semi-analytic model, 
or a limit on the duration from an ionization fraction of 5\% to 95\% of $\Delta^{90} z_{\rm re}  \limdelzwhkszpriordzamberpowerlawphys$ based on the AMBER simulations. 
}

\begin{document}

\title{\tittext}

\author[a]{P.~Chaubal,}
\author[b]{N.~Huang,\orcidlink{0000-0003-3595-0359}}
\author[a]{C.~L.~Reichardt,\orcidlink{0000-0003-2226-9169}}
\author[c,d,e]{A.~J.~Anderson,\orcidlink{0000-0002-4435-4623}}
\author[a]{B.~Ansarinejad,}
\author[e,d]{M.~Archipley,\orcidlink{0000-0002-0517-9842}}
\author[f]{L.~Balkenhol,\orcidlink{0000-0001-6899-1873}}
\author[g]{D.~R.~Barron,\orcidlink{0000-0002-1623-5651}}
\author[f]{K.~Benabed,}
\author[h,d,e]{A.~N.~Bender,\orcidlink{0000-0001-5868-0748}}
% ...

\affiliation[a]{School of Physics, University of Melbourne, Parkville, VIC 3010, Australia}
\affiliation[b]{Department of Physics, University of California, Berkeley, CA, 94720, USA}
\affiliation[c]{Fermi National Accelerator Laboratory, MS209, P.O. Box 500, Batavia, IL, 60510, USA}
\affiliation[d]{Kavli Institute for Cosmological Physics, University of Chicago, 5640 South Ellis Avenue, Chicago, IL, 60637, USA}
\affiliation[e]{Department of Astronomy and Astrophysics, University of Chicago, 5640 South Ellis Avenue, Chicago, IL, 60637, USA}
\affiliation[f]{Sorbonne Universit\'e, CNRS, UMR 7095, Institut d'Astrophysique de Paris, 98 bis bd Arago, 75014 Paris, France}
\affiliation[g]{Department of Physics and Astronomy, University of New Mexico, Albuquerque, NM, 87131, USA}
\affiliation[h]{High-Energy Physics Division, Argonne National Laboratory, 9700 South Cass Avenue, Lemont, IL, 60439, USA}
% ...

\abstract{\abstracttext{}}


\maketitle

\section{Introduction}
\label{sec_introduction}

At small angular scales, observations of the sky at millimeter wavelengths provide information about the secondary cosmic microwave background (CMB) anisotropies, radio galaxies, and the dusty, star-forming galaxies that comprise the cosmic infrared background (CIB). 
Multi-frequency observations can disentangle these components using their differing spectral and spatial dependencies, allowing us to constrain models for the different astrophysical signals simultaneously. 

The kinematic and thermal Sunyaev-Zel'dovich (SZ) effects are the largest secondary anisotropies at arcminute scales. 
Both SZ effects involve the scattering of CMB photons by free electrons along the photons' path. 
One result of this scattering is to induce a Doppler shift in the scattered photons, known as the kinematic SZ (kSZ) effect \citep{sunyaev72,sunyaev80}. 
The kSZ signal is proportional to $(v/c) n_e$ where $v$ is the bulk velocity of the electrons, and $n_e$ is the number density of free electrons. 
The kSZ signal is frequently separated into a contribution from the epoch of reionization (the patchy kSZ signal; e.g., \citenum{gruzinov98,knox98}), and a contribution from late-time large scale structure (the homogeneous kSZ signal; e.g., \citenum{shaw12,battaglia13}).

Inverse Compton scattering also yields a net transfer of energy from the (hotter) free electrons to the (colder) CMB photons, known as the thermal SZ (tSZ) effect \citep{sunyaev70,sunyaev72}. 
Unlike the kSZ effect, the tSZ effect distorts the CMB's initial blackbody spectrum, leading to a deficit of photons below $\sim$ 217\,GHz and an excess at higher frequencies. 
The amplitude of the tSZ signal scales as $(k_B T_e/m_ec^2)n_e$, where $m_e$ is the electron mass and $T_e$ is the temperature of the electrons.  
The tSZ anisotropy power spectrum depends steeply on the normalization of the matter power spectrum and thus can be used to place constraints on the growth of large-scale structure \citep[e.g.,][]{komatsu02}. 

In addition to the secondary CMB anisotropies, millimeter-wave maps also show clear evidence for emission from galaxies.
The main emission sources are synchrotron-dominated active galactic nuclei (AGN, e.g., \citenum{dezotti10}) and the cosmic infrared background (CIB) due to thermal dust emission from dusty, star-forming galaxies (DSFGs, e.g., \citenum{fixsen98,hauser01,lagache05,dole06}). 
The apparent flux of these galaxies is sometimes amplified by gravitational lensing \citep{hezaveh13}. 
While it is straightforward to mask the rare tail of very bright galaxies, it is impossible to exclude all emission from galaxies in this manner since the number of galaxies  in each instrumental beam is large. 
We can separate emission from radio galaxies and the CIB from the kSZ and tSZ effects using both angular and spectral information.

The angular power spectra of the millimeter sky have been measured previously at arcminute scales using data from the Atacama Cosmology Telescope (ACT; \citenum{ das14,louis25}), and the predecessor instruments to SPT-3G on the South Pole Telescope: SPT-SZ and SPTpol \citep{lueker10, shirokoff11, reichardt12b, george15, reichardt21}. 
These measurements have been used to constrain the level of tSZ and kSZ power using template-based weighted averages over angular scales near $\ell=3000$, with comparable results being found by both experiments.
The most recent ACT DR6 results have found  $\dtsz= 3.9\pm0.5\,\mu{\rm K}^2$ (at 143\,GHz) and $\dksz =2.0\pm0.9\,\uksq{}$ for the tSZ and kSZ power respectively, at  $\ell=3000$ in the nominal foreground model of that work   \citep{louis25, beringue25}. 
Previous SPT measurements have found $\dtsz=3.42\pm 0.54\,\uksq{}$ (at 143\,GHz)  and $\dksz =3.0\pm1.0\,\uksq{}$,  again at $\ell=3000$ \citep[][hereafter R21]{reichardt21}. 
These works have also studied the properties of the CIB and correlation between the CIB and galaxy clusters.

In this work, we present the first measurements using data from  the SPT-3G receiver of the high-$\ell$ ($\ell > 3000$) temperature power spectra. 
We use observations at 95, 150, and 220\,GHz of 1646\,\sqdeg{} of sky made during the 2019 and 2020 winters. 
These data allow us to measure the temperature anisotropy power in all six frequency cross-spectra at high signal-to-noise across angular multipoles from $1700 \le \ell \le 11,000$. 
Compared to the recent ACT DR6 measurement \citep{louis25}, this SPT-3G power spectrum measurement is more precise at $\ell \gtrsim 2900$ at 95\time95\,GHz, at $\ell \gtrsim 2800$ at 150\time150\,GHz, and at $\ell \ge 1700$ at 220\time220\,GHz.
The SPT-3G data significantly improve existing measurements of the temperature power in all six frequency cross-spectra at $\ell\gtrsim3000$. 
We use these data to  derive parameter constraints on the thermal and kinematic SZ effects, CIB, and radio galaxies.

The outline of this work is as follows. 
We describe the instrument and observations in  \S\ref{sec:datasets}. 
The power spectrum estimation framework is detailed in \S\ref{sec:estimator}, and 
the results of systematics tests performed are reviewed in \S\ref{sec:nulltests}. 
The resulting power spectra measurements are shown and discussed in \S\ref{sec:bandpowers}.
The expected emission sources at these observing frequencies and our modelling of this emission is discussed in \S\ref{sec:models}. 
We consider the fit quality in \S\ref{subsec:fitquality}. 
We present the parameter constraints in \S\ref{sec:params} before concluding in \S\ref{sec:conclusion}.


\section{Data}
\label{sec:datasets}

We use data collected by the SPT-3G receiver on the 10\,m South Pole Telescope during the austral winters of 2019 and 2020. 
This dataset, which has been labeled SPT-3G D1, is the same dataset 
as used by \cite{zebrowski25,camphuis25} to measure the primary CMB anisotropy power spectra at larger angular scales. 
The SPT-3G receiver consists of approximately 16,000 polarization-sensitive bolometers distributed in trichroic pixels at 95, 150, and 220\,GHz \citep{sobrin22}.
With an instantaneous field of view of 1.88$^\circ$ in diameter, SPT-3G is well-suited to conducting large-area surveys of the millimeter sky. 

This analysis uses the main survey region, covering the approximate RA range of 20$^\textrm{h}$28$^\textrm{m}$ to 3$^\textrm{h}$24$^\textrm{m}$ and declination range of $-41^\circ$ to $-71^\circ$. 
After applying the mask (see \S\ref{subsec:mask}), the effective sky area is  1646\,\sqdeg{}. 
The survey region is observed in four subfields, each of which covers the entire RA range but only one quarter of the declination range. 
Each subfield is observed by scanning the telescope left and then right across the full azimuthal range of the field, followed by a small elevation step, before repeating the process. 
 A full observation of any one of the four subfields takes roughly two hours. 
A total of 3280 observations pass data quality checks and are used in this analysis. 
The survey maps reach similar temperature noise levels in both years across the range $3000\le \ell \le 5000$.

\subsection{Map-making}
\label{sec:maps}

The maps used in this work are processed following the approach used by \cite{dutcher21},
and we refer the reader to that work for full details.
A paper with an in-depth discussion of the SPT-3G D1 maps is also in preparation, and expected to come out after this work. 
We summarize the approach here, focusing on the different parameter settings chosen to better recover smaller-scale modes. 
Briefly, the time-ordered data are band-pass filtered to remove atmospheric emission at low frequencies and to avoid aliasing high-frequency noise into the signal band. 
The high-pass filter is set at the angular frequency in the scan direction, $k_{\rm scan} > 500$, while the low-pass filter is at $k_{\rm scan} <$\, 15,000. 
We also remove the common mode across each detector wafer to further reduce large-scale atmospheric noise.  
Due to the scan orientation, for a Healpix\footnote{\url{https://healpix.sourceforge.io/}} map in celestial coordinates, these two filters are approximately a bandpass filter on the spherical harmonic coefficients $a_{\ell m}$ allowing $m \in $\,[500, 15,000]. 
   The data are then binned into Healpix $N_{\rm side}$ = 8192 pixels, with weights based on the mean power spectral density of the time-ordered data between frequencies corresponding roughly to $\ell \in [2000,4000]$. 
   
   To reduce the disk space and computational requirements of the analysis, we coadd data from multiple observations of all four subfields into a set of 200 `bundle' maps, with each bundle map covering the entire field. 
   This is done by splitting the data on each subfield into 200 temporally contiguous  periods, with the length of each period varying slightly to equalize the noise level in the resulting bundle maps. 
We choose to use the same time periods for all three frequencies. 
   While this choice reduces the uniformity of the noise across the set of bundle maps at each frequency, it simplifies the bookkeeping to avoid noise bias in the power spectrum estimator described next.

\section{Power spectrum estimation}
\label{sec:estimator}

As in previous SPT power spectrum analyses, we estimate the power spectra following the pseudo-Cl framework of \cite{hivon02} while using cross-spectra between data subsplits \citep{tristram05} to avoid issues of noise bias. 
We report bandpowers which are binned estimates of the power spectra, 
\be
D_\ell = \frac{\ell (\ell+1)}{2\pi} C_\ell,
\ee
with units of $\mu K_{\rm CMB}^2$.

We begin by calculating a biased, pseudo-Cl estimate  of the sky power, $\tilde{D}^{i\times j}_b$, for the frequency pair ($i \times j$) from the mean of the set of all cross-spectra:
\be
\tilde{D}^{i\times j}_b = \frac{1}{N_{\rm pair}}\sum_{p > q}  \frac{1}{2 N_b}\sum_{\ell\in b} \frac{\ell (\ell+1)}{2\pi}({a}^{i,*}_{p,\ell m} {a}^{j}_{q,\ell m} + {a}^{j,*}_{p,\ell m} {a}^{i}_{q,\ell m}),
\ee
where $a^{i,*}_{p,\ell m}$ is the spherical harmonic transform of the masked bundle map for frequency $i$ and time period $p$. 
$N_b$ is the number of modes within the bin $b$, and $N_{\rm pair} = 200 * 199 / 2$  is the number of unique, temporally disjoint pairs of bundle maps (the observations have been split into 200 temporal periods and coadded). 
This process yields a biased estimate of the true sky power as the maps have been filtered (see \S\ref{sec:maps}) to reduce instrumental and atmospheric noise and include the smoothing effect of the instrumental beams, among other effects.

Following \cite{hivon02}, we correct this biased estimate by estimating the information loss and mode-mixing due to the finite sky, and inverting the binned version of this operator. 
We describe the calculation of the mode-coupling matrix and transfer function in the following sections. 
The binned kernel $K^{i \times j}_{b b^\prime}$ for the frequency pair ($i \times j$) is:
\be \label{eqn:kernel}
K^{i \times j}_{b b^\prime} = P_{b\ell} \left(M_{\ell \ell^\prime}[\textbf{W}]\,F^{i\times j}_{\ell^\prime}B^{i}_{\ell^\prime}B^{j}_{\ell^\prime}\right)Q_{\ell^\prime b^\prime}.
\ee
Here $P_{b\ell}$ and $Q_{\ell^\prime b^\prime}$ are the binning operator and its reciprocal, $F^{i\times j}_{\ell^\prime}$ is the transfer function, and $B^{i}_{\ell}$ is the beam function for frequency $i$ along with the effects of the map pixel window function. 
An unbiased estimate of the binned sky is then calculated using the inverse of this kernel matrix as:
\be
D^{i\times j}_b = (K^{i \times j})^{-1}_{b b^\prime} \tilde{D}^{i\times j}_{b^\prime}
\ee

\subsection{Mask and mode-coupling matrix }\label{subsec:mask}

We determine the mask $\textbf{W}$ based on the geometric mean of the weight arrays of the 150\,GHz bundle maps. 
We threshold these weights at 40\% of the median geometric mean weight, and apodize within this region by a 15$^\prime$ Hanning window. 
We also mask emissive sources detected at greater than 6\,mJy in the SPT-3G 150\,GHz band. 
We choose not to mask massive galaxy clusters, and instead deal with the non-Gaussian sample variance due to these clusters when interpreting the observed thermal SZ power. 
For each source, the mask is set to zero out to a radius where the expected source flux would be detected at less than 1\,$\sigma$, beyond which the mask values taper smoothly from 0 to 1 according to a Gaussian function with $\sigma = 5^\prime$. 
The mode-coupling matrix $M_{\ell \ell^\prime}[\textbf{W}]$ for the mask $\textbf{W}$ is calculated using the public NaMaster software package \citep{alonso19}.\footnote{https://github.com/LSSTDESC/NaMaster}

\subsection{Simulations} 
\label{subsec:sims}

Simulations are used in the calculation of the transfer function ($F^{i\times j}_{\ell}$) and sample variance contribution to bandpower uncertainties. 
The input sky realizations are drawn as Gaussian realizations of a Planck best-fit cosmology plus secondary anisotropies and CIB taken from the \citetalias{reichardt21} best-fit model at the SPT-3G band centers.  
Radio galaxies are simulated as Poisson realizations of the de Zotti model \citep{dezotti10} with a single spectral index of $\alpha_{\rm rg} = -0.7$. 
These maps are convolved with the respective beam for each frequency band. 
For each simulated sky, we randomly choose two of the 200 bundles to simulate from the time-ordered data to maps. 
These two bundles are crossed to yield a cross-spectrum estimate of the filtered sky. 
This approach is taken to minimize the computational cost of running the end-to-end simulations, while remaining unbiased by effects due to incomplete coverage patterns. 
We do see extra variance in this approach compared to simulating the entire data set at $\ell\sim10,500$, especially in the 95\,GHz maps where the beam-convolved power is close to zero, however this variance is very small compared to the instrumental noise at those scales. 
We use 100 sky realizations in the transfer function and covariance calculations. 

\subsection{Covariance estimation and conditioning}
\label{sec:cov}

We construct a noisy estimate of the covariance using the scatter of the cross-spectra between the bundle maps and 100 noise-free sky simulations (see \citenum{lueker10}). 
Noise in this covariance estimate can affect model parameter inferences \citep{hartlap07,hamimeche09,taylor13,dodelson13, balkenhol22}. 
As in previous SPT power spectrum papers, we condition the noisy estimate to reduce the uncertainty in the covariance estimate. 
This covariance conditioning is described in the following paragraphs.

We expect most terms in the covariance to be nearly diagonal in $\ell$ since the native resolution of the survey field size is small compared to the $\Delta\ell=50$ binning used in estimating the covariance. 
Thus, we impose a requirement that the variances are uncorrelated beyond a separation of $\Delta \ell = 50$, i.e.~these terms are set to zero. 
We estimate the  average correlation at $\Delta \ell = 50$ by averaging the ratios of values on the diagonal to values one bin off the diagonal of the sample and noise covariance matrix. 
The assumption of no correlation is broken by one signal term: the Poisson sample variance due to radio galaxies in the simulations is highly correlated between angular multipoles. 
This term is estimated using the autospectrum of the 95\,GHz simulations (where it is the largest fraction of power) and scaled to the other frequency bands using the known frequency scaling used to make the simulations.  
For the covariance estimate, we also reduce the Poisson term due to radio galaxies by 15\% to approximately reproduce the radio power observed at 95 and 150\,GHz.

We estimate the diagonal signal variance terms from the set of 100 simulated sky realizations. 
We expect the diagonal of the variance of these simulations to follow the form:
\be
\textbf{C}_{bb}^{ij,kl} = \frac{1}{\nu_b} ( D_b^{ik} D_b^{jl} + D_b^{il} D_b^{jk} )
\ee
where $D_b^{ik}$ is the known theory spectrum for bin $b$ and frequency $i \times j$ cross-spectrum \citep{tristram05}. 
The prefactor, $\frac{1}{\nu_b}$ , where $\nu_b$ is an  effective number of degrees of freedom, will depend on $\ell$ and also the map filtering. 
For the sample variance terms, we fit the observed prefactors, $\frac{1}{\nu_b}$, to a function of the form $A + B \frac{1}{\ell-500}$ across the $\ell$-range of $\ell \in [1500, 9500]$, and use the results of the fit prefactors in the range $\ell \in [2000, 12,500]$ (the measured values are used at lower $\ell$). 
We use $\ell-500$ for the number of modes due to the high-pass filter that affects low $m$ modes. 
We also allow for extra variance in the angular multipole ranges where the simulations show increased variance due to the incomplete pixel coverage in the cross-spectrum of a single pair of simulated bundle maps. 
This should be conservative as the extra variance at these scales will be lower in the data cross-spectra which use all the bundle maps.   
The fit values are used at high $\ell$ due to the noisy covariance estimate at these scales.

We estimate the noise covariance terms using the observed scatter between bundle cross-spectra. 
We allow for instrumental noise that is correlated between frequency bands (due to atmospheric fluctuations)  at large angular scales ($\ell < 6000$). 
At higher multipoles, we see no evidence for such correlations and explicitly set these terms to zero. 
We estimate the noise auto-spectra terms at higher multipoles ($\ell > 6000$) from the autospectrum blocks, eg. $N_\ell^{150\times150\,{\rm GHz}}$ from the $150\times150\times150\times 150$\,GHz covariance block. 
In both cases, we smooth the logarithm of the diagonals with a smoothing scale of $\Delta \ell = 250$ (five bins) to reduce the scatter.

\subsection{Beams, bandpasses and calibration}
\label{sec:beams}

The instrumental beam function $B_\ell$ for each frequency band  is estimated using a combination of bright AGN in the field  and observations of Saturn.
We form a composite beam map using the field sources at small radii ($r \lesssim 1.\farcm5$) and Saturn at large radii ($r \gtrsim 3\farcm$).
The intermediate radii are constructed from a weighted mean of the field sources and Saturn, with the weight of the field source map falling linearly from 1 to 0 over $1.\farcm5 \lesssim r \lesssim 3\farcm$, while the weight of the planet maps increases from 0 to 1 over the same region. 
Before coadding, we account for multiplicative and additive offsets between the field sources and Saturn.
To account for the chromaticity of the beams, we create beams for several assumed spectral energy distributions (SEDs): a falling power law with $\alpha = -0.7$ for radio galaxies; a modified blackbody with $\beta = 1.8$ and $T = 25$\,K for the CIB; the derivative of the CMB blackbody for CMB fluctuations; and the non-relativistic tSZ SED.
We transform the beam between assumed SEDs using the physical model (introduced in \citenum{ge25,camphuis25}).
The physical model only reproduces the main lobe of the beam, so we assume that the sidelobes are frequency-independent to better agree with their observed changes between observing bands (we include an empirically determined term for sidelobe chromaticity in the beam uncertainty).
The reported bandpowers have been corrected by the estimated beams for a CMB black	body spectrum. 
When comparing the measured bandpowers to models, we adjust the predicted model spectra by the ratio of the beam functions estimated for the CMB and the beam for the closest SED for each component of the model, e.g., the modeled tSZ power spectrum is adjusted by the appropriate ratio of the CMB beam to the non-relativistic tSZ beam. 
A paper describing the details of the beam measurement and physical beam model is in preparation.

Uncertainties in the beam functions are reported as a beam correlation matrix $ \pmb{\rho}^{\rm beam}_{bb^\prime}$ that is multiplied by the binned theory spectra and added to the covariance matrix when calculating the likelihood of a given model. 
The additional covariance $\textbf{C}^{\rm beam}$ due to the beam can be computed as:
\begin{equation}
\textbf{C}^{\rm beam}_{bb^\prime} = \pmb{\rho}^{\rm beam}_{bb^\prime}D^{\rm theory}_{b}D^{\rm theory}_{b^\prime}.
\end{equation}

The bandpasses of SPT-3G were measured using a Fourier transform spectrometer (FTS). 
We estimate the calibration uncertainty on the FTS to be 0.3, 0.4, and 0.7\,GHz for the 95, 150 and 220 GHz bands respectively. 
We take the product of the measured bandpasses and estimated atmospheric transmission from the am model \citep{ammodel} in order to 
  calculate an effective band center for each of the potential signals: the thermal SZ effect, the CIB, and synchrotron sources. 
These band centers are used in the parameter fits to compare to the bandpowers which are reported in CMB temperature units.

The absolute calibration of the spectra is determined by comparing the power spectra of the maps to the best-fit \planck{} 2018 model \citep{planck18-6} plus a foreground power estimate.  
The angular multipoles used at 95 and 150\,GHz are $\ell \in [856,1810]$, while a narrower multipole range or $\ell\in[1144,1538]$ is used for 220\,GHz due to the relatively higher levels of foreground emission. 
The final uncertainties in temperature are [0.62\%, 0.62\%, 1.14\%] at [95, 150, 220] GHz.
Note that the uncertainties are correlated between the three bands, especially between 95 and 150\,GHz, due to the common sample variance.


\section{Systematics tests}
\label{sec:nulltests}

Following the last SPT power spectrum measurement at these angular scales \citep{reichardt21}, we perform two null tests to search for systematic errors within the dataset. 
In each null test, the data is split in half according to the quantity of interest, and then differenced. 
We conduct a cross-spectrum analysis of the differenced maps, and correct the resulting null spectra by the inverse kernel calculated as in the normal analysis to compare the null spectra to the observed on-sky power. 
The first null test splits the data in half temporally, the first half minus second half null test. 
The second null test splits the data in half according to the scan direction, the left minus right null test. 
We use an $\ell$-binning with a constant $\delta\ell=500$ so that different $\ell$-bins may be taken as uncorrelated.

Ideally, the null spectra would be consistent with zero. 
As the temperature spectra are measured at very high signal-to-noise, however, small changes can lead to failure, even if it does not bias the combined spectrum. 
For instance, if the calibration is slightly variable, a temporal null test might fail even if the mean calibration is accurately measured. 
To allow for such variations, we add a contribution equal to 10\% of the sample variance to the estimated error bars in the null tests. 
This level ensures any possible systematic effects will be small compared to the total covariance and have a statistically insignificant effect on later model fitting. 

The temporal split or first half minus second half null test passes in all three frequency bands. 
The PTEs of the measured $\chi^2$ are 10\%, 69\%, and 94\%. 
There is no evidence for a systematic bias in the first half minus second half null test at a level that would bias model fitting. 

As the D1 maps use time-ordered data that have not been deconvolved by the detector time constants nor the known timing offset between the pointing and detector data (see also \citenum{balkenhol23, camphuis25}), we expect and see residual power in the left minus right null test.
Before subtracting the expectation value due to the timing effects, we detect power in this null test in all frequency bands except 220\,GHz ($\chi^2 =$ 383.9, 102.6, and 13.3 respectively for 18 degrees of freedom (d.o.f.)).  
We compare the residual power to the predicted null power spectrum for an effective, not deconvolved, single-pole time constant, equal approximately to the difference between  the timing offset and the median measured time constant of the detectors, and we find good agreement. 

We  check the explanation that the excess is due to time constants on the 95\,GHz maps, which have the highest significance detection of residual power at $\chi^2 =$ 383.9. 
We remake the 95\,GHz maps, deconvolving the measured detector time constants and timing offset, and repeat the left minus right null test with these corrected maps. 
The left minus right null test on these corrected 95\,GHz maps is statistically consistent with zero ($\chi^2=18.0$, PTE = 46\%). 
Having confirmed the excess power in the left minus right null test can be removed by correcting for the time constants, we choose to proceed with the original maps as the instrumental beam measurement being used has been estimated for the non-deconvolved maps. 
The smoothing effect of the detector time constants and timing offset are captured in the beam estimate, as the beam is measured from bright AGN in the field. 
We conclude that, beyond the known effect of the detector time constants and time offsets,  the null tests show no evidence for unknown systematics that might significantly bias parameter constraints inferred from these data.

\begin{table*}
\begin{center}
\small
\begin{tabular}{cc|rr|rr|rr}
\hline\hline
\rule[-2mm]{0mm}{6mm}
& &\multicolumn{2}{c}{$95\,$GHz} &\multicolumn{2}{c}{$150\,$GHz} & \multicolumn{2}{c}{$220\,$GHz} \\
$\ell$ range&$\ell_{\rm eff}$ &$\hat{D}$ ($\mu{\rm K}^2$)& $\sigma$ ($\mu{\rm K}^2$) &$\hat{D}$ ($\mu{\rm K}^2$)& $\sigma$ ($\mu{\rm K}^2$)&$\hat{D}$ ($\mu{\rm K}^2$)& $\sigma$ ($\mu{\rm K}^2$) \\
\hline

1701 - 1800 & 1751.7 & 402.5 & 5.3 & 399.1 & 5.4 & 457.3 & 6.8\\
1801 - 1900 & 1853.0 & 314.9 & 4.1 & 310.3 & 4.2 & 367.9 & 5.8\\
1901 - 2000 & 1954.8 & 246.2 & 3.5 & 241.2 & 3.4 & 304.5 & 4.7\\
2001 - 2100 & 2051.8 & 235.8 & 3.2 & 232.5 & 3.2 & 305.7 & 4.6\\
2101 - 2200 & 2154.7 & 184.9 & 2.4 & 182.3 & 2.4 & 258.3 & 3.9\\
2201 - 2300 & 2252.9 & 138.7 & 1.8 & 135.1 & 1.8 & 214.9 & 3.2\\
2301 - 2400 & 2351.5 & 125.5 & 1.6 & 122.8 & 1.5 & 210.7 & 3.1\\
2401 - 2500 & 2453.3 & 105.4 & 1.3 & 102.4 & 1.3 & 193.9 & 2.9\\
2501 - 2700 & 2608.1 & 81.32 & 0.72 & 78.87 & 0.68 & 179.55 & 1.83\\
2701 - 3000 & 2873.2 & 55.55 & 0.42 & 53.01 & 0.37 & 167.10 & 1.38\\
3001 - 3300 & 3163.5 & 40.95 & 0.33 & 38.37 & 0.26 & 169.73 & 1.33\\
3301 - 3600 & 3458.0 & 32.85 & 0.30 & 30.98 & 0.21 & 182.68 & 1.38\\
3601 - 4000 & 3802.0 & 30.06 & 0.29 & 28.76 & 0.17 & 204.78 & 1.31\\
4001 - 4400 & 4198.8 & 30.13 & 0.33 & 29.48 & 0.17 & 238.75 & 1.47\\
4401 - 4800 & 4593.9 & 31.60 & 0.39 & 31.38 & 0.19 & 271.45 & 1.66\\
4801 - 5300 & 5042.8 & 33.86 & 0.44 & 34.87 & 0.19 & 316.70 & 1.68\\
5301 - 5800 & 5537.3 & 37.61 & 0.53 & 39.22 & 0.21 & 368.70 & 1.97\\
5801 - 6400 & 6083.0 & 42.03 & 0.63 & 45.22 & 0.23 & 432.50 & 2.09\\
6401 - 7000 & 6681.1 & 46.79 & 0.78 & 52.62 & 0.28 & 511.95 & 2.51\\
7001 - 7600 & 7281.1 & 53.06 & 0.99 & 60.01 & 0.33 & 592.18 & 2.91\\
7601 - 8300 & 7931.1 & 60.02 & 1.20 & 69.37 & 0.37 & 683.90 & 3.30\\
8301 - 9000 & 8626.6 & 67.47 & 1.60 & 79.93 & 0.48 & 794.75 & 4.06\\
9001 - 10000 & 9442.7 & 76.10 & 2.18 & 92.55 & 0.52 & 935.50 & 4.35\\
10001 - 11000 & 10430.4 & 85.99 & 3.79 & 108.80 & 0.74 & 1113.98 & 5.99\\
\comment{ 
2001 -  2200 &  2077   & 218.4 &    3.8   & 215.6 &    2.3   & 286.2 &    6.5   \\ 
2201 -  2500 &  2332   & 128.2 &    1.9   & 125.9 &    1.1   & 201.7 &    4.3   \\ 
2501 -  2800 &  2636   &  81.9 &    1.1   &  80.29 &   0.67   & 170.4 &    4.1   \\ 
2801 -  3100 &  2940   &  52.84 &   0.79   &  51.88 &   0.46   & 156.9 &    4.0   \\ 
3101 -  3500 &  3293   &  36.87 &   0.58   &  36.89 &   0.31   & 155.4 &    3.7   \\ 
3501 -  3900 &  3696   &  31.35 &   0.57   &  31.19 &   0.29   & 182.8 &    4.4   \\ 
3901 -  4400 &  4148   &  28.85 &   0.65   &  31.24 &   0.29   & 202.0 &    4.8   \\ 
4401 -  4900 &  4651   &  30.25 &   0.89   &  33.62 &   0.35   & 245.7 &    6.0   \\ 
4901 -  5500 &  5203   &  35.3 &    1.1   &  39.73 &   0.42   & 290.2 &    7.0   \\ 
5501 -  6200 &  5855   &  43.4 &    1.9   &  46.34 &   0.53   & 349.8 &    8.7   \\ 
6201 -  7000 &  6607   &  44.6 &    3.2   &  57.24 &   0.72   &   435. &    11.   \\ 
7001 -  7800 &  7408   &  46.7 &    6.7   &  69.5 &    1.2   &   524. &    15.   \\ 
7801 -  8800 &  8310   &    61. &    12.   &  89.0 &    1.8   &   665. &    21.   \\ 
8801 -  9800 &  9311   & - & -   &  98.7 &    2.9   &   729. &    34.   \\ 
9801 - 11000 & 10413   & - & -   & 122.0 &    4.5   &   962. &    49.   \\ 
}

\hline
\end{tabular}
\caption{\label{tab:bps1} 
Angular multipole range, weighted multipole value $\ell_{\rm eff}$, bandpower $\hat{D}$, 
and bandpower uncertainty $\sigma$ for the auto-spectra of the $95\,$GHz, $150\,$GHz, and $220\,$GHz maps with point sources detected at $>6$\,mJy at 150 GHz masked at all frequencies.
The uncertainties in the table are calculated from the diagonal elements of the covariance matrix, which includes noise and sample variance, but not beam or calibration errors. 
}
\normalsize
\end{center}
\end{table*}

\begin{table*}[htbp]
\begin{center}
\small
\begin{tabular}{cc|rr|rr|rr}
\hline\hline
\rule[-2mm]{0mm}{6mm}

& & \multicolumn{2}{c}{$95\times150\,$GHz} & \multicolumn{2}{c}{$95\times220\,$GHz} & \multicolumn{2}{c}{$150\times220\,$GHz} \\
$\ell$ range&$\ell_{\rm eff}$ &$\hat{D}$ ($\mu{\rm K}^2$)& $\sigma$ ($\mu{\rm K}^2$) &$\hat{D}$ ($\mu{\rm K}^2$)& $\sigma$ ($\mu{\rm K}^2$)&$\hat{D}$ ($\mu{\rm K}^2$)& $\sigma$ ($\mu{\rm K}^2$) \\
\hline
1701 - 1800 & 1751.7 & 398.2 & 5.3 & 394.8 & 5.7 & 408.1 & 5.9\\
1801 - 1900 & 1853.0 & 309.9 & 4.1 & 304.3 & 4.6 & 317.7 & 4.7\\
1901 - 2000 & 1954.8 & 240.9 & 3.5 & 237.1 & 3.7 & 250.7 & 3.8\\
2001 - 2100 & 2051.8 & 231.2 & 3.1 & 230.0 & 3.5 & 245.6 & 3.6\\
2101 - 2200 & 2154.7 & 180.5 & 2.4 & 179.7 & 2.7 & 196.0 & 2.8\\
2201 - 2300 & 2252.9 & 133.5 & 1.7 & 130.5 & 2.1 & 148.4 & 2.2\\
2301 - 2400 & 2351.5 & 120.7 & 1.5 & 119.8 & 1.9 & 138.6 & 2.0\\
2401 - 2500 & 2453.3 & 100.2 & 1.3 & 99.6 & 1.6 & 119.4 & 1.7\\
2501 - 2700 & 2608.1 & 76.24 & 0.66 & 76.67 & 0.92 & 97.67 & 0.99\\
2701 - 3000 & 2873.2 & 49.97 & 0.35 & 51.64 & 0.57 & 74.98 & 0.62\\
3001 - 3300 & 3163.5 & 34.84 & 0.25 & 36.77 & 0.46 & 63.35 & 0.51\\
3301 - 3600 & 3458.0 & 26.64 & 0.20 & 30.74 & 0.42 & 60.60 & 0.46\\
3601 - 4000 & 3802.0 & 23.22 & 0.17 & 28.42 & 0.38 & 63.72 & 0.40\\
4001 - 4400 & 4198.8 & 22.54 & 0.18 & 30.88 & 0.41 & 72.08 & 0.42\\
4401 - 4800 & 4593.9 & 22.86 & 0.20 & 32.98 & 0.46 & 80.71 & 0.46\\
4801 - 5300 & 5042.8 & 24.83 & 0.21 & 37.91 & 0.47 & 93.28 & 0.47\\
5301 - 5800 & 5537.3 & 27.30 & 0.25 & 44.88 & 0.55 & 108.58 & 0.53\\
5801 - 6400 & 6083.0 & 30.93 & 0.29 & 52.97 & 0.62 & 127.99 & 0.55\\
6401 - 7000 & 6681.1 & 34.77 & 0.35 & 62.45 & 0.76 & 151.76 & 0.66\\
7001 - 7600 & 7281.1 & 39.38 & 0.43 & 73.13 & 0.97 & 174.94 & 0.79\\
7601 - 8300 & 7931.1 & 44.5 & 0.5 & 85.3 & 1.2 & 202.7 & 0.8\\
8301 - 9000 & 8626.6 & 50.9 & 0.7 & 99.2 & 1.6 & 236.8 & 1.0\\
9001 - 10000 & 9442.7 & 59.2 & 0.8 & 118.2 & 2.0 & 276.3 & 1.1\\
10001 - 11000 & 10430.4 & 65.5 & 1.3 & 133.8 & 3.2 & 330.1 & 1.5\\

\hline
\end{tabular}
\caption{\label{tab:bps2} Angular multipole range, weighted multipole value $\ell_{\rm eff}$, bandpower $\hat{D}$, 
and bandpower uncertainty $\sigma$ for the cross-spectra of the $95\,$GHz, $150\,$GHz, and $220\,$GHz maps with point sources detected at $>6$\,mJy at 150 GHz masked at all frequencies.
The uncertainties in the table are calculated from the diagonal elements of the covariance matrix, which includes noise and sample variance, but not beam or calibration errors. 
}
\normalsize
\end{center}
\end{table*}



\section{Bandpowers}
\label{sec:bandpowers}

\begin{figure*}[t]\centering
\includegraphics[width=0.9\textwidth]{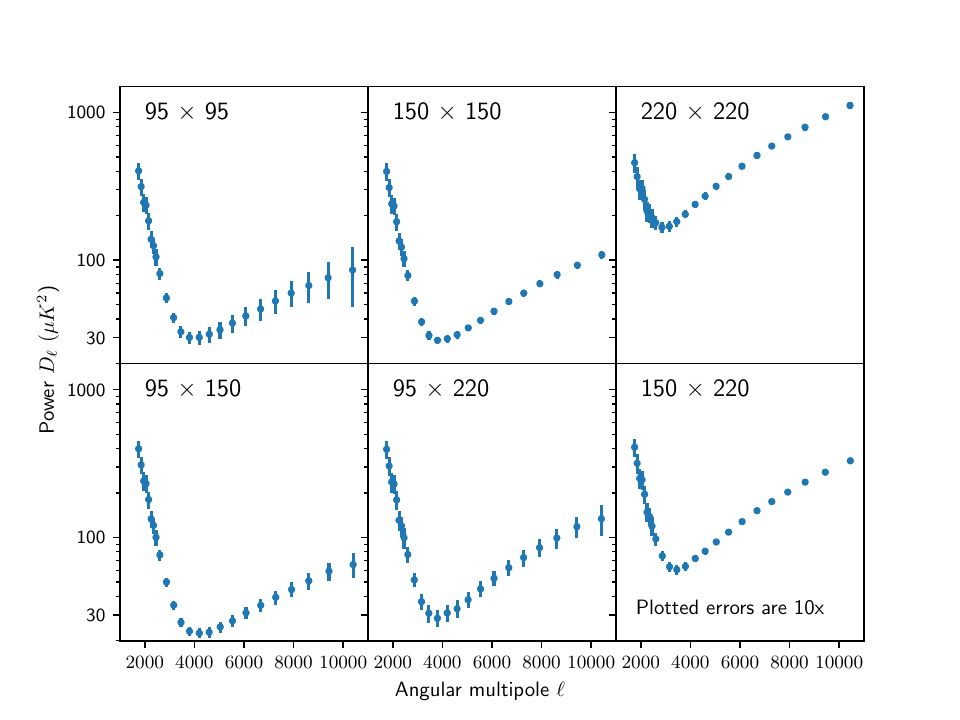}
\caption{  
The six auto- and cross-spectra measured with the 95, 150, and 220\,GHz SPT-3G data. 
To be more visible, the plotted uncertainties have been scaled up by a factor of ten. 
The plotted uncertainties are based on the diagonal elements of the covariance matrix, which includes noise and sample variance but not beam or calibration errors. 
  }
  \label{fig:bps}
\end{figure*}

The measured bandpowers from 1646\,\sqdeg{} of SPT-3G survey are plotted in Fig.~\ref{fig:bps} and tabulated in Tables \ref{tab:bps1} and \ref{tab:bps2}. 
Power is detected at very high S/N in all six frequency cross-spectra across angular multipoles in the range $1700 \le \ell \le 11,000$. 
The bandpowers and associated data products (such as the covariance matrix and bandpower window functions) required to compare the observations to theory spectra can be downloaded from the SPT and LAMBDA websites. 
The release includes likelihood codes that have been used with the cobaya MCMC sampler \citep{torrado21} to derive the model constraints reported in \S\ref{sec:params}.

The bandpower uncertainties for these SPT-3G data are significantly smaller than those for  previous  SPT measurements at these angular scales using data from SPT-SZ and SPTpol \citepalias{reichardt21}. 
Around $\ell \sim 3000$, the uncertainties have been reduced by factors of 2.0, 1.4, and 2.9 for the $95\times95$\,GHz, $150\times150$\,GHz and $220\times220$\,GHz power spectra respectively. 
The measurement is more precise than the ACT DR6 power spectrum measurement \citep{louis25} at all measured scales for $220\times220$\,GHz, and at $\ell > 2800-2900$ for $95\times95$\,GHz and $150\times150$\,GHz. 
The transition from sample-variance-dominated to noise-variance-dominated bandpower uncertainties is at $\ell \sim 4000,~5500$, and 5000 for the $95\times95$\,GHz, $150\times150$\,GHz, and $220\times220$\,GHz power spectra respectively, indicating that the measurement is sample variance dominated to fairly small angular scales. 
These maps and the resulting  bandpowers represent a significant improvement in our knowledge of the millimeter sky at arcminute scales.

\begin{figure*}[t]\centering
\includegraphics[width=0.9\textwidth]{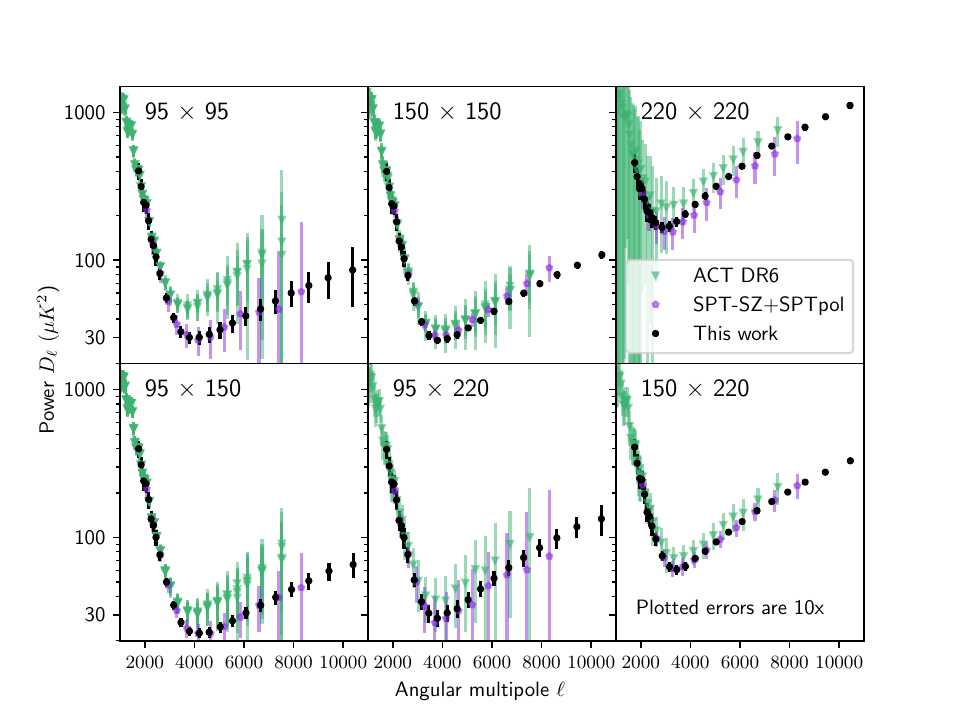}
\caption{  
The six auto- and cross-spectra measured with the 95, 150, and 220\,GHz SPT-3G data (black circles) are plotted along with measurements by ACT DR6 (light green, upside-down triangles, \citenum{louis25}) and a combination of SPT-SZ and SPTpol (purple pentagons, \citetalias{reichardt21}). 
Note that the plotted ACT bandpowers are array cross spectra and so have multiple points at the same $\ell$ for some frequency combinations. 
We stress that no correction has been made for the different observing bands or point source masking thresholds between the three measurements, which explains the observed differences in the power observed from AGN and dusty galaxies between the three experiments. 
ACT DR6 masked sources above 15 mJy at 150 GHz, while this work and \citetalias{reichardt21} masked sources above 6 mJy at 150 GHz. 
The plotted uncertainties are based on the diagonal elements of the covariance matrix, which include noise and sample variance, but not beam or calibration errors. 
Note that uncertainties for all three datasets have been increased by a factor of ten to be seen more easily. 
  }
  \label{fig:comparison}
\end{figure*}

We compare the new measurement against previous measurements by ACT \citep{louis25} and SPT-SZ/SPTpol \citepalias{reichardt21} in Fig.~\ref{fig:comparison}, with the uncertainties for all three experiments increased by a factor of ten to be more visibly apparent in the figure. 
Note that the effective frequency bandcenters and point source masking thresholds differ slightly between the three measurements, so we do not expect an exact match for the measured sky power. 
While all three mask bright point sources based on their 150\,GHz flux, the  masking threshold is 6\,mJy in the two SPT measurements and 15\,mJy in the ACT analysis. 
Thus one would expect (as is seen) more radio Poisson power in the ACT measurement. 
We have made no adjustment for how the bandcenters and source masking threshold would change the expected power from the tSZ effect, radio galaxies, or CIB between these measurements. 
Nonetheless, we can see qualitative agreement between the three measurements modulo the different levels of power from AGN and dusty galaxies.

\section{Modeling of millimeter-wave emission}
\label{sec:models}

The measured temperature anisotropy power spectra, shown in Fig.~\ref{fig:bps}, are expected to be sourced from a combination of the lensed primary CMB anisotropy, the thermal and kinematic SZ effects, CIB, correlations between the tSZ and CIB, a Poisson distribution of radio galaxies, and galactic cirrus. 
We review the models considered for each term in the following subsections. 
The fitting results, including the inferred power spectra for the different options, are discussed in \S\ref{sec:params}.

\subsection{Primary cosmic microwave background}\label{subsec:cmb}

The lensed primary CMB anisotropy is the largest contribution to the observed power at larger angular scales ($\ell \lesssim 2500$). 
We model the CMB power spectrum using CosmoPower \citep{spuriomancini22}. 
In all parameter fits, we allow the six \lcdm{} parameters to vary. 
Unless specifically noted, 
we include a multi-dimensional Gaussian prior taken from the Planck 2018 \lcdm{} chain for `plikHM\_TTTEEE\_lowl\_lowE' \citep{planck18-6} to inform the values of the six \lcdm{} parameters. 
While discussing the kSZ constraints, we comment in \S\ref{subsec:kszcmb} on the parameter shifts observed when using priors based on the `CMB-SPA' or `CMB-SPA no tau prior' chains from \cite{camphuis25}. 
As one would expect, the small-angular scale data of this work do not shift the posteriors of the six \lcdm{} parameters away from the prior  for any of the cases considered.

\subsection{Thermal Sunyaev-Zel'dovich effect}\label{subsec:tsz}

We consider three approaches to fitting the tSZ power.
\begin{enumerate}
\item \textbf{Template models:} 
The first approach follows past SPT works, and models the tSZ power using a single parameter that scales a simulation-based template for the tSZ power spectrum. 
\be  \label{eqn:tsztemplate}
D_\ell^{\rm tSZ,~143\,GHz} = \dtsz{} \left(\frac{D_{\ell}^{\rm tSZ ~simulation} }{D_{3000}^{\rm tSZ ~simulation}}\right),
\ee
where $\dtsz$ is a free parameter representing the tSZ power at $\ell=3000$ and 143 GHz  and $D_{\ell}^{\rm simulation}$ is one of the two simulation-based templates mentioned below. 
We assume a non-relativistic tSZ frequency scaling to go from 143\,GHz to the tSZ power at other frequencies. 
We have tested instead using a relativistic tSZ frequency scaling with $T \in [0,10]$ keV, and have found the data do not constrain the cluster temperature; the posterior is consistent with the input prior on $T$. 
We consider tSZ templates from two simulations. 
The first template is the Shaw tSZ model \citep{shaw10b}, and the second template is taken from Agora simulations  \citep{omori24}. 
In later discussions, we will refer to these as part of the G15 and Agora SZ template sets respectively. 
Both tSZ templates are plotted in figures in \S\ref{sec:params}. 

\item \textbf{$\ell^{\alpha}$ SZ models:}
The second approach  follows \cite{louis25} by introducing an angular-multipole-dependent rescaling of the simulation templates: 
\be\label{eqn:tszpowerlaw}
D_\ell^{\rm tSZ,~143\,GHz} = \dtsz{}  \left(\frac{\ell}{3000}\right)^{\alpha_{\rm tSZ}} \left(\frac{D_{\ell}^{\rm tSZ ~simulation} }{D_{3000}^{\rm tSZ ~simulation}}\right). 
\ee
These fits have two free tSZ parameters: an amplitude at $\ell=3000$ and 143\,GHz, $\dtsz{}$, and the exponent, $\alpha_{\rm tSZ}$. 
An appropriately chosen exponent $\alpha_{\rm tSZ}$ produces a near perfect match between the  Shaw and Agora tSZ templates across the measured angular scales, so this extra degree of freedom effectively eliminates the shape difference between the two tSZ templates considered.
We refer to these fits as `power law' or `$\ell^{\alpha}$' SZ fits, and use a similar power law treatment for both the kSZ and tSZ spectra.

\item \textbf{Free SZ model:}
Finally, we look at more complex variations by introducing a multipole-dependent scaling of the Shaw tSZ template, which we label `free SZ' in later sections. 
Based on an exploration of where the data constrain the tSZ power, we use seven independent points across the angular multipole range to fit the SPT-3G data.
The set is  $a^{\rm tSZ}_{\ell=0} = a^{\rm tSZ}_{\ell=1700}, ~a^{\rm tSZ}_{\ell=2300},$ $a^{\rm tSZ}_{\ell=3000},$ $a^{\rm tSZ}_{\ell=4000},$ $a^{\rm tSZ}_{\ell=5000},$ $a^{\rm tSZ}_{\ell=6000},$ $a^{\rm tSZ}_{\ell=8000}=a^{\rm tSZ}_{\ell=16000}$. 
We note that while we are rescaling a template, the choice between scaling the Agora or Shaw tSZ templates is unimportant given the number of points in rescaling. 
We calculate the rescaling factor $f_{\rm tSZ}(\ell)$ as a function of angular multipole $\ell$ using the monotonic cubic spline function \texttt{scipy.interpolate.PchipInterpolator}, and multiply the simulation-based tSZ template by this scale-dependent rescaling factor: 
\be\label{eqn:tszspline}
D_\ell^{\rm tSZ,143\,GHZ} = f_{\rm tSZ}(\ell)  \left(\frac{D_{\ell}^{\rm tSZ ~simulation} }{D_{3000}^{\rm tSZ ~simulation}}\right).
\ee
We refer to results for these fits as `free SZ', as we simultaneously use a monotonic cubic spline approach for both the tSZ and kSZ power spectra.

\end{enumerate}

\subsection{Kinematic Sunyaev-Zel'dovich effect}\label{subsec:ksz}

Paralleling the modeling of the tSZ effect, we consider three approaches to quantify the kSZ power in these data. 

\begin{enumerate}
\item \textbf{Template models:}
First, we parameterize the kSZ power by a single amplitude parameter $\dksz$ for the kSZ power level at $\ell=3000$ that rescales a simulation-based prediction for the power spectrum $D_{\ell}^{\rm kSZ ~simulation}$:
\be \label{eqn:ksztemplate}
D_\ell^{\rm kSZ} = \dksz{} \left(\frac{D_{\ell}^{\rm kSZ ~simulation} }{D_{3000}^{\rm kSZ ~simulation}}\right).
\ee
We use two templates. 
The first, as in \citetalias{george15} and \citetalias{reichardt21}, is constructed by  setting the power of the cooling and star formation (CSF) homogeneous kSZ template from \cite{shaw12} and patchy kSZ template from \citep[][hereafter Z12]{zahn12} to be equal at $\ell=3000$.
 We label fits using this combined Shaw$+$Zahn kSZ template as `G15 SZ' in plot legends. 
The second template is taken from the kSZ power spectrum of the Agora simulations  \citep{omori24}, and only includes the homogenous kSZ power spectrum (i.e. contributions from the late-time universe). 
Both kSZ templates are plotted in figures in \S\ref{sec:params}. 

\item \textbf{$\ell^\alpha$ SZ models:}
Second, as with the tSZ modeling, we allow the simulated kSZ template to be rescaled by a factor of $\ell^{\alpha_{\rm kSZ}}$. 
\be\label{eqn:kszpowerlaw}
D_\ell^{\rm kSZ} = \dksz{}  \left(\frac{\ell}{3000}\right)^{\alpha_{\rm kSZ}} \left(\frac{D_{\ell}^{\rm kSZ~ simulation} }{D_{3000}^{\rm kSZ ~simulation}}\right). 
\ee
Marginalizing over the exponent $\alpha_{\rm kSZ}$ is able to capture most of the differences between the two simulation templates. 
As noted in \S\ref{subsec:tsz}, we introduce a power law scaling for both the tSZ and kSZ spectra. 
We refer to results from these chains as the `$\ell^\alpha$ G15 SZ' or `$\ell^\alpha$ Agora SZ' results depending on the simulation template used. 

\item \textbf{Free SZ model:}
In this model, we scale the G15 kSZ template by a multipole-dependent factor, $f_{\rm kSZ}(\ell)$, using a monotonic cubic spline function. 
We use the following points: 
$a^{\rm kSZ}_{\ell=0} = a^{\rm kSZ}_{\ell=2300}, ~a^{\rm kSZ}_{\ell=3000},~a^{\rm kSZ}_{\ell=4000},$ $a^{\rm kSZ}_{\ell=5000},~a^{\rm kSZ}_{\ell=6000},$ $a^{\rm kSZ}_{\ell=8000},$ $~a^{\rm kSZ}_{\ell=10000}=a^{\rm kSZ}_{\ell=16000}$. 
Unlike the tSZ, we do not use a point at $\ell=1700$ for the kSZ as the kSZ power is highly degenerate with the primary CMB power. 
As in the tSZ case, the kSZ template choice is not important. 
The modeled kSZ power spectrum is then:
\be \label{eqn:kszspline}
D_\ell^{\rm kSZ} = f_{\rm kSZ}(\ell)  \left(\frac{D_{\ell}^{\rm kSZ ~simulation} }{D_{3000}^{\rm kSZ ~simulation}}\right).
\ee
We label these results `free SZ', noting that we use a similar rescaling of both the kSZ and tSZ power spectrum. 
\end{enumerate}

\subsection{Cosmic infrared background}\label{subsec:modelcib}

We model the CIB by a Poisson term with constant $C_\ell$ plus a clustering term. 
We assume the Poisson power has a modified blackbody spectrum, $S_\nu \propto \nu^{\beta_P} B_\nu(T)$, with grey-body index $\beta_P$. 
The temperature $T$ is assumed to be 25\,K for all chains. 
Thus, the Poisson CIB power is described by two parameters: the grey-body index $\beta_P$ and the power at $\ell=3000$ and 150\,GHz, $D_{3000}^{\rm CIB~pois,\,150\,GHz}$.

We consider two models for the clustering CIB power.
\begin{enumerate}
\item \textbf{Template model:} 
This model follows previous SPT works \citepalias[e.g.,][]{george15,reichardt21} and breaks the clustering power into one-halo and two-halo clustering templates taken from the best-fit halo model in \cite{viero13a}. 
This clustering model has two free parameters related to its shape: the amplitude parameters for the one-halo and two-halo clustering templates at $\ell=3000$ and 150\,GHz, and is expressed as:
\bea \label{eqn:cibtemplate}
D_\ell^{\rm CIB, cl} &=& D_{3000}^{\rm 1-halo} \left(\frac{D_{\ell}^{\rm 1-halo ~simulation} }{D_{3000}^{\rm 1-halo ~simulation}}\right) +\nonumber \\
&&D_{3000}^{\rm 2-halo} \left(\frac{D_{\ell}^{\rm 2-halo ~simulation} }{D_{3000}^{\rm 2-halo ~simulation}}\right).
\eea
Similar to the Poisson CIB, the spectral dependence is taken to follow  a modified blackbody spectrum $S_\nu \propto \nu^{\beta_C} B_\nu(T)$, with the grey-body index $\beta_C$ allowed to float. 
The temperature is fixed  to $T=25$\,K.

\item \textbf{Free CIB model:} The second model uses the measurements to reconstruct the shape of the CIB clustering power at 150\,GHz. 
The CIB clustering power at 150\,GHz is taken to be the product of a rescaling factor $f_{\rm CIB}(\ell)$ times  the approximate clustering shape $D_\ell\propto \ell^{0.8}$ used in previous work \citep[e.g.,][]{lueker10} to estimate the clustering power:
\be \label{eqn:cibspline}
D_\ell^{\rm CIB~cl,150\,GHZ} = f_{\rm CIB}(\ell)  \left(\frac{\ell}{3000}\right)^{0.8}.
\ee
The rescaling factor $f_{\rm CIB}(\ell)$ is calculated with the monotonic cubic spline function with five free parameters (three more than the template case):
$a^{\rm CIB~cl, 150\,GHz}_{\ell=0} = a^{\rm CIB~cl, 150\,GHz}_{\ell=1800}$, $a^{\rm CIB~cl, 150\,GHz}_{\ell=3000}$, $a^{\rm CIB~cl, 150\,GHz}_{\ell=4200}$,  $a^{\rm CIB~cl, 150\,GHz}_{\ell=6000}$,  $a^{\rm CIB~cl, 150\,GHz}_{\ell=9,000} = a^{\rm CIB~cl, 150\,GHz}_{\ell=16,000}$. 
The frequency dependence of the CIB power is again described by a modified black-body SED, however 
the grey-body index $\beta_C(\ell)$ is now multipole-dependent and is calculated with a monotonic cubic spline function, using three degrees of freedom. 
The parameters are $\beta_C(0) = \beta_C(2000)$, $\beta_C(5000)$, $\beta_C(9000) = \beta_C(16000)$. 
Uniform priors from 0 to 4 are assumed for all three grey-body indices. 
In total, this model uses eight free parameters to describe the clustered CIB power. 
Note we also use a monotonic cubic spline for the tSZ-CIB correlation in this case, as discussed in \S\ref{subsec:tszcib}. 
This `free CIB' approach has the advantage of being agnostic to the choice of CIB model templates.

\end{enumerate}

\subsection{Galactic cirrus}
We use two approaches to model thermal dust emission from our own Milky Way. 
\begin{enumerate}
\item \textbf{Template models:}
In the template fits, we use the cirrus model from \citetalias{george15}, with an updated prior on the power in the autospectrum at each frequency at $\ell=80$ based on a cross-correlation of the SPT-3G survey data with the \textit{Planck} satellite \cite[see][]{camphuis25}. 
\item \textbf{Free CIB model:}
When we use the free CIB model for clustered CIB power, we assume the CIB term includes Galactic dust given its similar spectral dependence and do not use a separate cirrus term nor cirrus prior. 
\end{enumerate}

\subsection{Correlation between the thermal Sunyaev-Zel'dovich effect and cosmic infrared background}\label{subsec:tszcib}

The galaxies of the CIB and the hot plasma that sources the tSZ effect are both tracers of the same large-scale distribution of matter, and therefore we expect a positive spatial correlation between the CIB and tSZ signals. 
We model the signal due to the tSZ-CIB correlation in the frequency cross-spectrum for $\nu_1 \times \nu_2$ as:
\bea\label{eqn:tszcib}
D^{\rm tSZ-CIB,~\nu_1 \times\nu_2}_\ell &=& -\xi(\ell) \left( \sqrt{D^{\rm CIB, \nu_1 \times \nu_1}_{\ell}  D^{\rm tSZ, \nu_2 \times \nu_2}_{\ell} } \right. \nonumber \\
&&\left. + \sqrt{D^{\rm CIB, \nu_2 \times \nu_2}_{\ell} D^{\rm tSZ, \nu_1 \times \nu_1}_{\ell}}    \right). 
\eea
We assume the tSZ-CIB correlation $\xi(\ell)$ is a function of angular multipole $\ell$ only and require $\xi(\ell)\ge 0$.

As with the tSZ and CIB modeling, we consider two approaches, one based on templates from simulations and one based on using a monotonic cubic spline to describe the $\ell$-dependence of the tSZ-CIB correlation. 
\begin{enumerate}
\item \textbf{Template models:}
In the template case,  the angular dependence is taken from a simulation, with a floating amplitude parameter $ \xi_{3000}$  rescaling the correlation function:
\be \label{eqn:tszcibtemplate}
 \xi(\ell) = \xi_{3000} \left( \frac{\xi^{\rm simulation}(\ell)}{\xi^{\rm simulation}(3000)} \right). 
\ee
We have run chains for two tSZ-CIB template choices. 
As in \citetalias{reichardt21}, the first template for the correlation is the form found by \citetalias{zahn12}, when looking at the CIB simulations of \cite{shang12}. 
This template rises with angular multipole. 
The second template we use is taken from the Agora simulations \citep{omori24}. 
In these simulations the tSZ-CIB correlation peaks around $\ell\sim2000$ and falls towards zero at high angular multipoles. 
Chains using these templates for the tSZ-CIB correlation also use the 1- and 2-halo templates for the CIB power spectra. 

\item \textbf{Free CIB model:}
As an alternative to the template fits, we use a monotonic cubic spline function to calculate $ \xi(\ell)$ based on parameter values for the correlation at specified angular multipoles. 
This approach uses $\xi(1700)$, $\xi(2100)$, $\xi(2500)$, $\xi(3000)$, $\xi(3500)$, $\xi(4200)$, $\xi(5000)$, $\xi(6000)$,  and $ \xi(0)  =  \xi(16000)  = 0$  for eight free parameters. 
Note that chains using  this form of the tSZ-CIB correlation also use the monotonic cubic spline for the CIB clustering shape and grey-body index, and results from these chains are labeled `free CIB'.
\end{enumerate}

\subsection{Radio galaxies} \label{subsec:modelrg}

We model the population of radio galaxies below the masking threshold of 6\,mJy at 150\,GHz as a Poisson distribution. 
We assume the SED of each radio galaxy follows a power law $S_\nu \propto \nu^{\alpha_{\rm rg}}$ where the exponent is drawn from a normal distribution $N(\alpha_{\rm rg},\sigma^2_{\rm rg})$. 
The fits are largely unaffected by the choice to allow  $\sigma^2_{\rm rg}\ge0$, although there is a small improvement in the fit quality with the template fits ($\Delta \chi^2 \sim 3$ for 1 d.o.f.). 
Normalizing the power by $D_{3000}^{\rm rg,~150\,GHz}$ at $\ell=3000$ and 150\,GHz, the modeled power for a Poisson distribution of radio sources at 150\,GHz is:
\be\label{eqn:rg}
D^{\rm 150\times150}_\ell =D_{3000}^{\rm rg,~150\,GHz} \left(\frac{\ell}{3000}\right)^2
\ee
In all chains, we allow the Poisson power level, $D_{3000}^{\rm rg,~150\,GHz} $, mean spectral index, $\alpha_{\rm rg}$, and variance, $\sigma^2_{\rm rg}$,  to float freely within uniform priors for three free radio parameters.

\section{Quality of fit }
\label{subsec:fitquality}
\begin{table*}[ht]
\begin{center}
\small
\begin{tabular}{cc|c| c| c}
\hline\hline
\rule[-2mm]{0mm}{6mm}
Full model:  & &Agora & free CIB, $\ell^\alpha$ G15 SZ& free CIB+SZ\\
\hline
&$\chi^2$& 177.5 & 140.4 & 129.7  \\
&$N_{\rm model}$ & 11 &  25 & 35\\
& PTE (\%) & 0.6 &  8.8 & 8.6\\
\hline
\hline
term & d.o.f. &\multirow{2}*{\delchisq\ } &\multirow{2}*{\delchisq\ }  & \multirow{2}*{\delchisq\ } \\
 removed & removed &   &  &  \\
\hline
Radio Poisson& 3 & +2089.& +1416.  & +1450.\\  
DSFG Clustering &3/8/8&  +227.& +358.& +333.\\
DSFG Poisson&2 & +44.& +287. &+334.\\ 
tSZ (and tSZ-CIB) & 2/10/15 &  +944. &	+747. & +725.\\ 
tSZ-CIB & 1/8/8 &+20.  & +14. & +14.\\
kSZ & 1/2/7 &  +30. & +5. & +14.\\ 
\end{tabular}
\caption{\label{tab:chisq} 
The top row has the best-fit $\chi^2$ for three models, while the second row has the number of model parameters used to describe the emission from the tSZ and kSZ effects, radio galaxies, CIB, and correlations between these terms. 
The left column has numbers for the template fit using Agora templates for the kSZ, tSZ, and tSZ-CIB. 
While not shown, the minimum $\chi^2$ for the template fit using the G15 templates is similar to the  $\chi^2$  of the Agora templates.  
The middle column has the results for  the CIB-free and powerlaw SZ fit, while the right column has the numbers for the CIB+SZ-free model. 
The bottom section shows the increase in best-fit $\chi^2$  as components are removed from the model, measured relative to the $\chi^2$ for the complete model. 
Note that setting the tSZ power to zero also zeros the tSZ-CIB correlation. 
We note that in several cases other parameters hit  prior bounds as terms are removed, and thus the prior bounds affect the increase in $\chi^2$. 
We list a single number for the d.o.f.~when all model options have the same number of parameters, and list three numbers for the respective cases when the number of degrees of freedom changes between the models. 
The overall fit quality is poor, especially for the template fits, although all individual model terms lead to significant improvements in the quality of fit. 
} \normalsize
\end{center}
\end{table*}

In Tab.~\ref{tab:chisq}, we list the quality of the fits for the three models that are the focus of the next section:  the Agora template model;  the free CIB with $\ell^\alpha$ SZ model; and the CIB+SZ free model.  
We give the increase in the best-fit $\chi^2$ as model components are removed (zeroed) from the fit.  Regardless of model, the data require including all the components with the exception of Galactic cirrus where the posteriors remain prior-driven.
The quality of the fit rapidly declines when the CIB, radio galaxy, or tSZ terms are removed, with  increase to the \chisq{} in the hundreds to thousands. 
There are also significant reductions in the fit quality when the tSZ-CIB correlation is set to zero or the kSZ power is set to zero ($\delchisq\sim 15$). 
While not shown, zeroing the primary CMB power also dramatically increases the \chisq{} as one would expect. 
Modeling the measured bandpowers requires all of the terms described in the last section.

\begin{figure}[t]\centering
\includegraphics[width=0.7\textwidth]{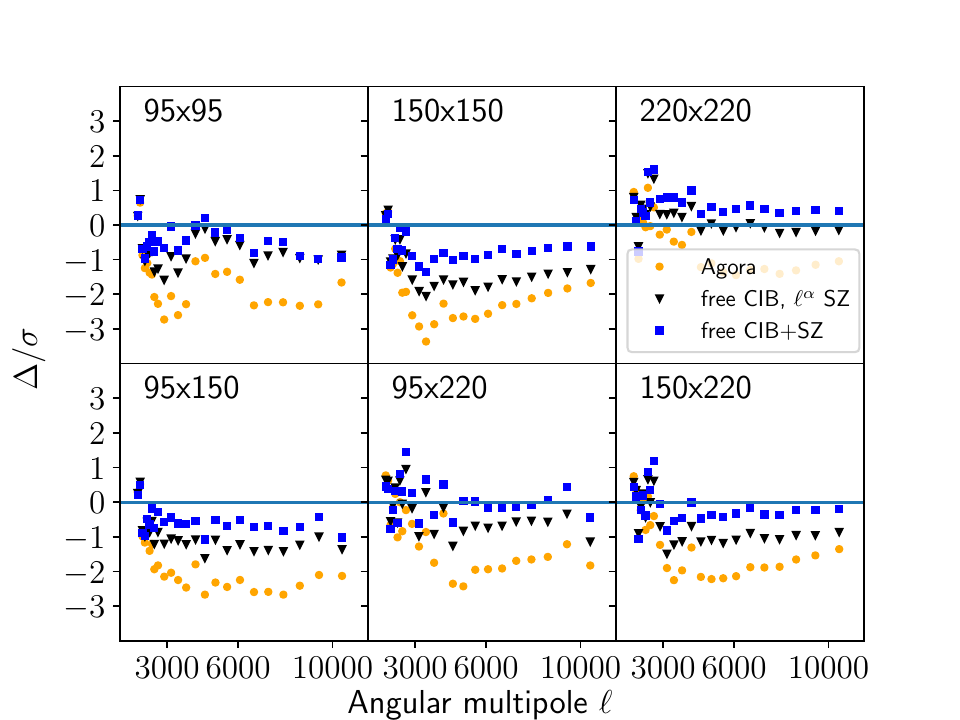}
\caption{  
To illustrate where the data and best-fit models diverge, we show the bandpower differences between the model and data divided by the square root of the diagonal of the covariance matrix. 
The three cases shown are the best-fit points of the three models in Table~\ref{tab:chisq}. 
  }
  \label{fig:residuals}
\end{figure}

We highlight that we achieve only a borderline quality of fit, based on the best-fit $\chisq{}$ relative to the d.o.f.~in the data minus the d.o.f.~in the model. 
The PTE is implausibly low for the template models, and borderline at 8-9\% for the other two models. 
For the latter two cases, this may be an indication that the monotonic cubic spline approach is inefficient in terms of d.o.f. 
For instance, with respect to the CIB clustering term, this approach needs five parameters to achieve the same \chisq{} as are achieved with two amplitude parameters for the 1- and 2-halo CIB clustering templates. 
However, this cannot be the full explanation, if only because it does not explain the poor fit to the template models. 
The residuals for the three best-fit models, plotted in terms of $\frac{\Delta D_b }{ C_{bb}^{0.5}}$, are shown in Fig.~\ref{fig:residuals}. 

We put forward three potential explanations for the modest fit quality. 
First, this could reflect an underestimate of the covariance matrix, in particular, the beam uncertainty. 
Doubling the beam uncertainties can provide a reasonable PTE for all three models (10s of percent), although we stress that we believe this would be a significant over-estimate of the beam uncertainty. 
Second, the low fit quality could be due to a mis-estimate in the instrument model, such as the beam measurement, FTS band measurement,  or modeling of the transfer function of the maps. 
Third, the model may be missing an important component,  over-simplifying the model of one or more of the emission sources, or use an inaccurate SED for one or more components. 
We have tried model extensions such as the clustering of radio galaxies, correlations between the tSZ and radio power, or correlations between the CIB and radio galaxies. 
We find these extensions can improve the fit quality, but only for unphysical values that are ruled out by other measurements. 
We have also tried other extensions, such as CO line emission, which did not improve the fit quality.

 \begin{table*}[ht]
\begin{center}
\small
\begin{tabular}{l||c |c |c | c  c|}
\rule[-2mm]{0mm}{6mm}
\multirow{2}*{Label} & \multirow{2}*{tSZ model} & \multirow{2}*{kSZ model} & \multirow{2}*{tSZ-CIB model} & \multicolumn{2}{c|}{Clustered CIB}\\
& & & & $\ell$-shape & SED\\
\hline
\hline
\textbf{Agora} & Eqn.~\ref{eqn:tsztemplate}  & Eqn.~\ref{eqn:ksztemplate}  & Eqn.~\ref{eqn:tszcibtemplate}  & Eqn.~\ref{eqn:cibtemplate} & $\beta$ \\
\textbf{templates}& Agora template & Agora  template & Agora template & & \\
\hline
\textbf{free CIB, }& Eqn.~\ref{eqn:tszpowerlaw}  & Eqn.~\ref{eqn:kszpowerlaw}   & $\xi(\ell)$&  Eqn.~\ref{eqn:cibspline} & $\beta(\ell)$ \\
\textbf{$\ell^\alpha$ G15 SZ}& Shaw template& Shaw+Zahn template &(spline)&&(spline)\\
\hline
\textbf{free CIB+SZ}& Eqn.~\ref{eqn:tszspline}  & Eqn.~\ref{eqn:kszspline} & $\xi(\ell)$&  Eqn.~\ref{eqn:cibspline} & $\beta(\ell)$ \\
& (spline) & (spline) & (spline) & & (spline) \\
\hline
\hline
G15 templates & Eqn.~\ref{eqn:tsztemplate}  & Eqn.~\ref{eqn:ksztemplate}  & Eqn.~\ref{eqn:tszcibtemplate}  & Eqn.~\ref{eqn:cibtemplate} & $\beta$ \\
& Shaw template &  Shaw+Zahn template  & Shang template & & \\
\hline
free CIB, & Eqn.~\ref{eqn:tsztemplate} & Eqn.~\ref{eqn:ksztemplate}   & $\xi(\ell)$ &  Eqn.~\ref{eqn:cibspline} & $\beta(\ell)$ \\
G15 SZ& Shaw  template& Shaw+Zahn  template & (spline) & &(spline)  \\
\hline
free CIB, & Eqn.~\ref{eqn:tsztemplate} & Eqn.~\ref{eqn:ksztemplate}   & $\xi(\ell)$ &  Eqn.~\ref{eqn:cibspline} & $\beta(\ell)$ \\
Agora SZ& Agora  template & Agora  template & (spline) & &(spline)  \\
\hline
free CIB,& Eqn.~\ref{eqn:tszpowerlaw}  & Eqn.~\ref{eqn:kszpowerlaw}   & $\xi(\ell)$&  Eqn.~\ref{eqn:cibspline} & $\beta(\ell)$ \\
 $\ell^\alpha$ Agora SZ& Agora  template & Agora template&(spline) &&(spline) \\
\hline
\end{tabular}
\caption{\label{tab:models} 
A breakdown of the different model choices discussed in \S\ref{sec:params}. 
Results under the top three models are shown in Fig. \ref{fig:szspectra}, \ref{fig:tszcibpower}, and \ref{fig:cibpower}.
A monotonic cubic spline is used for the tSZ-CIB correlation and modified black body index, see \S\ref{subsec:tszcib} and \S\ref{subsec:modelcib}, when marked by $\xi(\ell)$ and $\beta(\ell)$ respectively. 
Terms for the Poisson distributions of radio galaxies (\S\ref{subsec:modelrg}) and DSFGs (\S\ref{subsec:modelcib}) are not listed, as the same model is used for these terms in all cases. 
} \normalsize
\end{center}
\end{table*}

 \begin{figure*}[t]\centering
\includegraphics[width=0.45\textwidth]{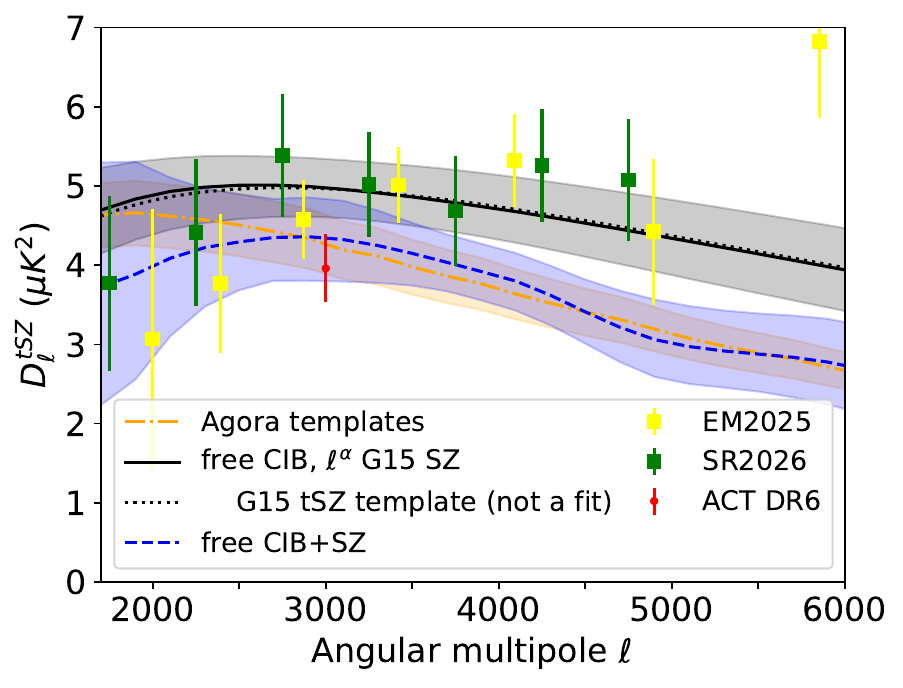}
\includegraphics[width=0.45\textwidth]{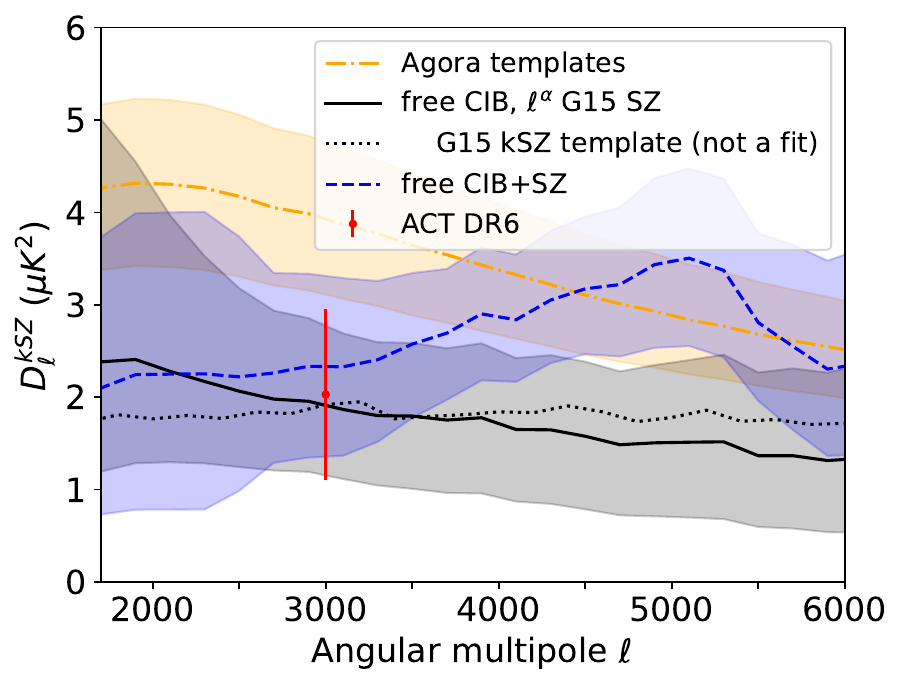}
\caption{  \label{fig:szspectra}   
Results for the median power (lines) and 68.3\% confidence region (shaded area) for the tSZ power (\textbf{left panel}) and kSZ power (\textbf{right panel}). 
The orange dot-dashed lines and region mark the results when using the Agora templates for the tSZ, kSZ, and tSZ-CIB spectra, along with a CIB fit for the Poisson, 1-halo, and 2-halo terms. 
The other lines all use the free CIB model. 
The black solid line and grey region show the results for the $\ell^\alpha$ rescaling of the SZ templates used by G15. 
We show the unscaled shape of the G15 kSZ and tSZ templates with dotted lines; we stress that the amplitudes are set to match the black line at $\ell=3000$, rather than a fit. 
The $\ell^\alpha$ rescaling favors a tSZ power spectrum very similar to the G15 tSZ predictions, but a kSZ power spectrum that falls towards small angular scales as predicted by Agora. 
The blue dashed line and region show the inferred SZ power in the free CIB+SZ  case. 
We note that the inferred SZ power is fairly stable across model choices near $\ell=3000$, but we see variations at smaller angular scales, primarily in the better-constrained tSZ power spectrum. 
For reference, we show in red the constraints from Planck + ACT DR6 \lcdm{} chains released with \citenum{louis25}. 
The DR6 results agree well with our results at $\ell=3000$. 
We show the tSZ constraints from \cite{efstathiou25} in yellow and from \cite{raghunathan26} in green, which except for the  highest multipole bin of each one are within 2\,$\sigma$ of all three fits of this work.
  }
\end{figure*}
 
 \begin{figure*}[t]\centering
\includegraphics[width=0.9\textwidth]{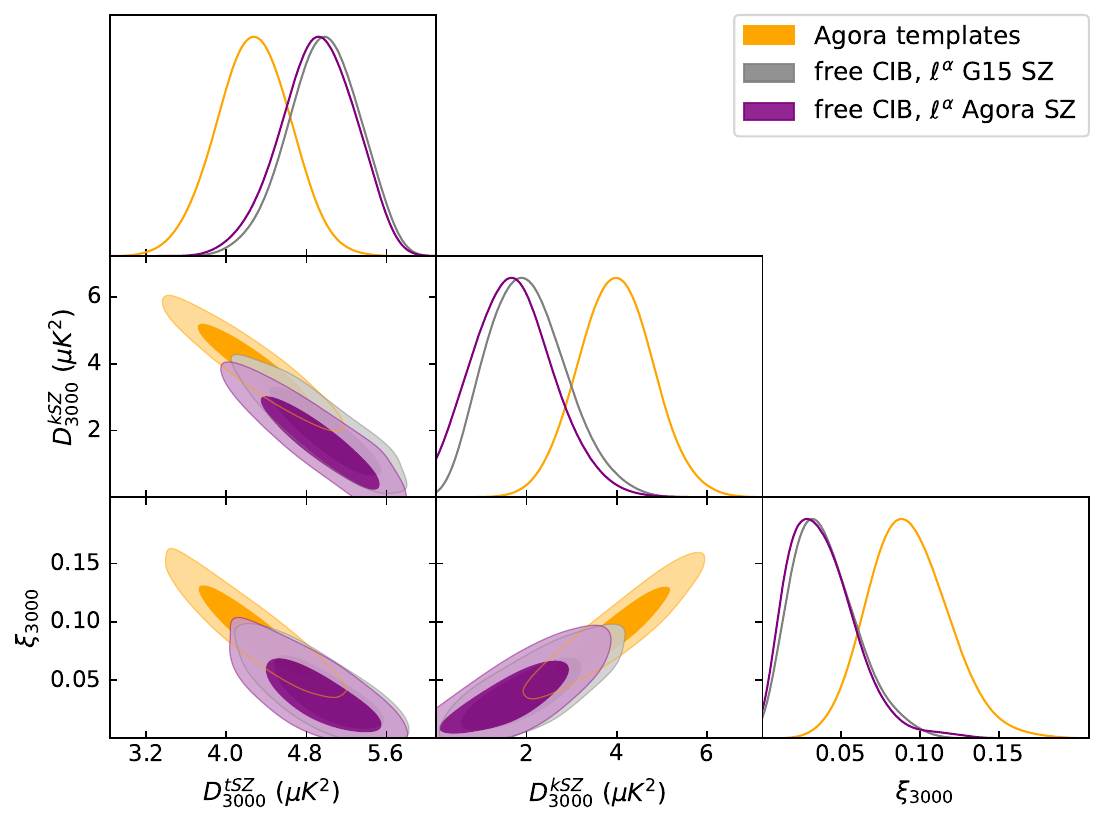}
\caption{  
1D and 2D posteriors for the kSZ power (\dksz), tSZ power at 143\,GHz (\dtsz), and tSZ-CIB correlation ($\xi_{3000}$), all evaluated at $\ell=3000$. 
The results when using a template fit based on the Agora tSZ, kSZ, and tSZ-CIB spectra are shown in orange, while the results for the more flexible free CIB model with an $\ell^\alpha$ rescaling of the SZ templates is shown in grey for the G15 kSZ and tSZ templates and in purple for the Agora kSZ and tSZ templates. 
The inferences are robust between these cases, with no significant shifts between the three model choices. 
There are clear degeneracies between all three parameters with higher levels of tSZ power correlating to lower levels of kSZ power and tSZ-CIB correlation. 
Equivalently, higher correlation between the tSZ and CIB emission reduces the tSZ power and increases the kSZ power. 
  }
  \label{fig:SZpower}
\end{figure*}

\section{Parameter constraints}
\label{sec:params}

We now look at the constraints placed on the secondary CMB anisotropies, radio galaxies, and CIB by the SPT-3G power spectra measurement, 
for the models described in \S\ref{sec:models}. 
We determine the posteriors on model parameters using Markov chain Monte Carlos (MCMCs) run with the \texttt{cobaya} package  \citep{torrado21}.
As noted in \S\ref{sec:bandpowers}, the data release includes the likelihood code used for these MCMCs. 
All chains include the SPT-3G data and a prior on the \lcdm{} model parameters.
Except for a discussion of the fits' CMB dependence in \S\ref{subsec:kszcmb}, the \lcdm{} prior is based on the results reported by \cite{planck18-6}.  

We focus our discussion of the parameter posteriors for the top three models in bold in Table~\ref{tab:models} that have also been discussed in \S\ref{subsec:fitquality}, though we also mention at times other model choices in that table to build intuition about how the posteriors depend on the modeling assumptions. 
These three models are the Agora template fits; the free CIB, $\ell^\alpha$ G15 SZ fits; and the free CIB+SZ fits. 

 \begin{figure}[t]\centering
\includegraphics[width=0.7\textwidth]{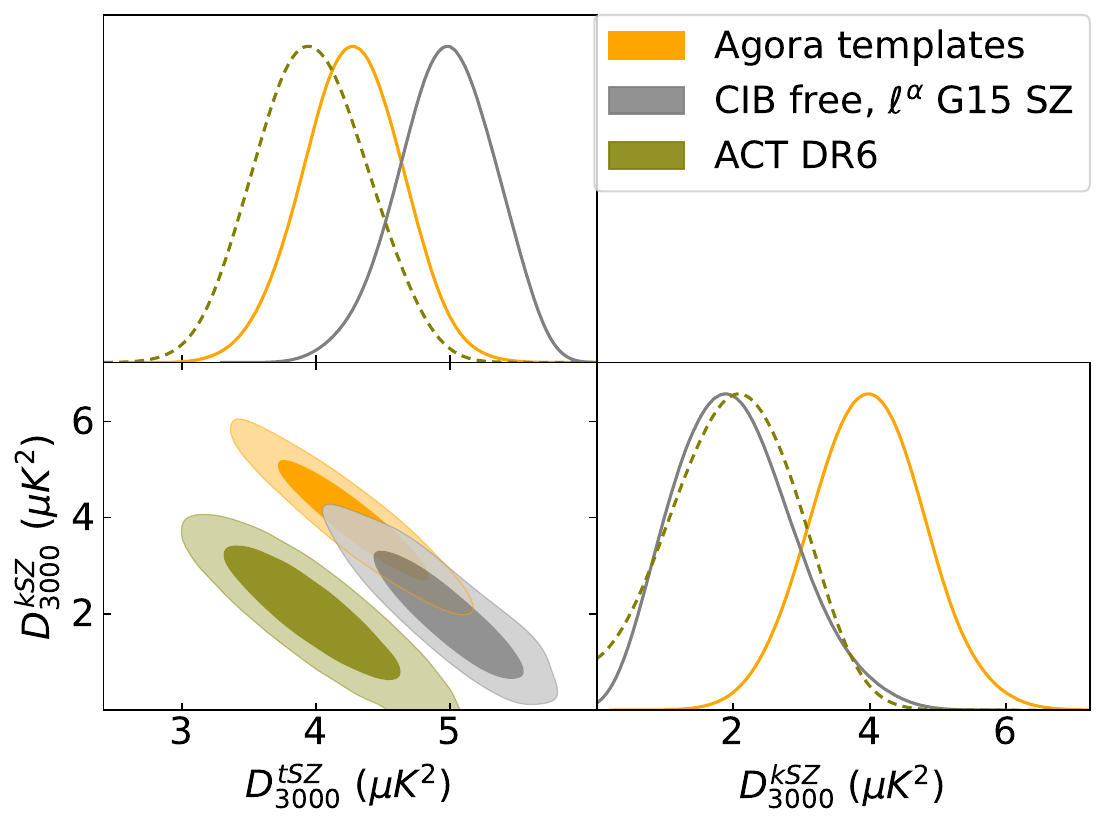}
\caption{  
1D and 2D posteriors for the kSZ power (\dksz) and tSZ power at 143\,GHz (\dtsz{}), both evaluated at $\ell=3000$. 
The results when using a template fit based on the Agora tSZ, kSZ, and tSZ-CIB spectra are shown in orange, while the results for the more flexible free CIB model with an $\ell^\alpha$ rescaling of the SZ templates is shown in grey  for the G15 kSZ and tSZ templates. 
Recent ACT+ \textit{Planck}  results  \citep{louis25}  are shown in olive, with dashed lines for the 1D posteriors. 
The SPT-3G data sharpen our knowledge of the kSZ and tSZ power. 
  }
  \label{fig:szact}
\end{figure}

\subsection{Constraints on SZ power}\label{subsec:sz}

At $\ell=3000$, we detect positive tSZ power $\sim$ 4.5\,\uksq{} at 143\,GHz for the three major model choices (and most of the alternatives), while
the kSZ power shows larger variations ($2-4\,\uksq$) between the models. 
The tSZ and kSZ power constraints as a function of angular multipoles for these three models are plotted in Fig.~\ref{fig:szspectra} and tabulated at select angular multipoles in Table~\ref{tab:sz}. 
We remind the reader that the plotted posteriors on the tSZ power do not include the expected non-Gaussian tSZ sample variance, and the non-Gaussian sample variance should be included when testing a model against these posteriors.

Notably compared to previous SPT works, two of these three models offer flexibility in the angular dependence of the SZ power spectra and tSZ-CIB spectra. 
One argument for this flexibility is the improved quality of fit for these models, as reported earlier in \S\ref{subsec:fitquality}. 
However another reason to allow flexibility  is that we observe significant template-dependence on the SZ power levels with the current data, with parameter shifts of up to $\sim5\,\sigma$ between the template choices. 
Using the same model (including kSZ, tSZ, tSZ-CIB, and 1- and 2-halo clustered CIB templates) as \citetalias{reichardt21}, we find the tSZ power at $\ell=3000$ and 143 GHz to be  $\dtsz = \dtszbase$ and the frequency-independent kSZ power at $\ell=3000$ to be  $\dksz \limkszbase$ at 95\% confidence. 
Using the same CIB model but replacing the  templates used by \citetalias{george15} and \citetalias{reichardt21}  for the tSZ and kSZ power spectra and tSZ-CIB correlation by templates  derived from the Agora simulations, 
we recover $\dtsz = \dtszagora$ and $\dksz = \dkszagora$. 
Clearly despite the apparent precision of the results for each set of templates, the specific allocation of power between the tSZ and kSZ effects is highly dependent on the exact form of the templates assumed. 

We explore which templates matter the most to these shifts by swapping one template at a time, and find the different templates assumed for the tSZ-CIB correlation explain about 60\% of the parameter shifts. 
Specifically, when still using the Agora templates for the kSZ and tSZ effect while swapping the  tSZ-CIB correlation shape from the Agora form to the \citetalias{zahn12} form used by \citetalias{george15}, the SZ fits shift from $\dtsz = \dtszagora$ to $\dtszagorawbasetszcib$  and from $ \dksz = \dkszagora$ to $\dkszagorawbasetszcib$. 
The tSZ template accounts for most of the remaining parameter shifts.
The equivalent swap from the Agora tSZ power spectrum template to the Shaw tSZ template used by G15 yields 
$\dtsz = \dtszagorawbasetsz$  and $ \dksz =\dkszagorawbasetsz$. 
Changing the kSZ template alone leads to small ($\sim$\,$0.5\,\sigma$) changes in the inferred tSZ and kSZ power at $\ell=3000$. 
The large parameter shifts between different simulation templates strongly motivate the exploration of the non-template fits for the SZ power and tSZ-CIB correlation.

\subsubsection{SZ constraints with the free CIB model}

Given the demonstrated dependence on the tSZ-CIB modeling,  we now turn to the `free CIB' chains where monotonic cubic spline functions are used to recover the CIB clustering power and tSZ-CIB correlation. 
We will consider the results when using templates for the kSZ and tSZ power, before looking at the results when allowing the additional flexibility of the power law SZ fits. 
The CIB+SZ free case will be discussed in \S\ref{subsec:cibszfree}. 

As noted in the last paragraph, differences between the simulation predictions for the tSZ-CIB correlation have a significant effect on the derived SZ powers. 
With this more flexible tSZ-CIB (and CIB) modeling but still using templates for the kSZ and tSZ power spectra, we find $\dtsz = \dtszfreecibphysbasesz$ and $\dksz =  \dkszfreecibphysbasesz$ for the tSZ and kSZ templates used by G15 and R21. 
This shifts to $\dtsz = \dtszfreecibphysagorasz$ and $\dksz =\dkszfreecibphysagorasz$ for the Agora tSZ and kSZ templates. 
These results are much closer to the pure template fits using the Agora SZ and tSZ-CIB templates rather than G15 templates.  
Investigating, we note that the tSZ-CIB shape recovered in these cases is a much better match to the Agora prediction than the \citetalias{zahn12} prediction for the tSZ-CIB shape as will be discussed further in \S\ref{subsec:szcibparams}.

The remaining differences between two sets of SZ templates largely disappear when we move to the $\ell^\alpha$ SZ case (still using the `free CIB' model for the CIB and tSZ-CIB terms). 
This is illustrated in Fig.~\ref{fig:SZpower}, which has a triangle plot of the kSZ power at $\ell=3000$, the tSZ power at 143\,GHz and $\ell=3000$, and the  tSZ-CIB correlation at $\ell=3000$ for the Agora template fits and the two power law SZ fits. 
For both the kSZ and tSZ power spectra, the $\ell^\alpha$ factor is able to account for nearly all the template differences over the angular scales constrained by the data. 
As should be expected to account for the template differences, the best-fit exponents $\alpha_{\rm tSZ}$ and $\alpha_{\rm kSZ}$ are template-dependent. 
There is only a  $0.3\,\sigma$ shift for the tSZ power and $0.5\,\sigma$ in the kSZ power when moving from G15 to Agora tSZ and kSZ templates, as can be seen in Fig.~\ref{fig:SZpower}. 
Therefore, for the rest of this work, we will only show numbers for the $\ell^\alpha$ rescaling of the G15 tSZ and kSZ power spectrum templates, and will not discuss the equivalent rescaling of the Agora SZ templates. 
These results are labeled in figures as `free CIB, $\ell^\alpha$ G15 SZ'. 
For this case, the posterior on the kSZ power is $\dksz = \dkszpowerlawphys$ while the posterior for the tSZ power is $\dtsz = \dtszpowerlawphys$. 

We compare the posteriors on the kSZ and tSZ power from these data and the ACT DR6 chains in Fig.~\ref{fig:szact}. 
Neither work has masked galaxy clusters. 
Both datasets show a similar degeneracy between the kSZ and tSZ power (where an increase in kSZ power can be balanced by a decrease in tSZ power). 
While the ACT and SPT-3G 1D posteriors for both the kSZ and tSZ power agree reasonably due to this degeneracy, the SPT-3G data prefer more overall SZ power at $\ell=3000$, especially in the Agora template fits. 
An investigation of the agreement at other multipoles and of whether these shifts are related to the different foreground models between these analyses is beyond the scope of this work.

As shown by the black line in the left panel of Fig.~\ref{fig:szspectra}, the recovered tSZ power in the G15 power-law SZ fit is very similar to the Shaw tSZ template and peaks at $\ell\sim 2600$. 
This result suggests a tSZ spectrum peaking at smaller angular scales than then  the best-fit tSZ spectrum for the ACT DR6 power spectra \citep{louis25}, which peaks at $\ell\sim 1500$ and falls towards higher $\ell$. 
The power law fit for the kSZ power falls slightly across these angular scales,  as shown by the black line in the right panel of the figure. 
The recovered shape is consistent with the kSZ templates from either G15 or Agora. 
Note that these power-law exponents are sensitive to the predicted CMB power in the range $1700\le \ell \le 3000$, as will be discussed in \S\ref{subsec:kszcmb}.

\subsubsection{SZ constraints with the free CIB+SZ model}\label{subsec:cibszfree}

Finally, we look at the `free CIB+SZ' results when monotonic cubic spline functions are used in the modeling of the SZ power, as well as the CIB  and tSZ-CIB correlation. 
These results are intended to clarify where the constraining power of the data lies, as well as to serve as a template-agnostic test of what the data can tell us about the SZ power spectra. 
The free CIB+SZ case is shown in blue in Fig.~\ref{fig:szspectra}, with tSZ results in the left panel, and kSZ results in the right panel. 
The SZ power levels at selected $\ell$s are also reported in Tab.~\ref{tab:sz}. 
The inferred tSZ power is consistent at $\lesssim 1\,\sigma$ with the Agora template fit across this angular multipole range. 
The inferred tSZ power spectrum peaks around $\ell\sim2900$. 
The uncertainties on tSZ power become large below $\ell=2500$ (due to CMB sample variance) and at $\ell>6000$ due to instrumental noise, high levels of CIB power, and uncertainty in the tSZ-CIB correlation.

\begin{table*}[ht]
\begin{center}
\scriptsize
\begin{tabular}{c||c|c |c  c c |}
\rule[-2mm]{0mm}{6mm}
Model &$D_{2000}^{\rm tSZ}$&$D_{3000}^{\rm tSZ}$&$D_{4000}^{\rm tSZ}$&$D_{5000}^{\rm tSZ}$&$D_{6000}^{\rm tSZ}$\\
\hline
Agora templates & -& \dtszagora &- & - &- \\ 
free CIB, $\ell^\alpha$ G15 SZ& \dtsztwokpowerlawphys & \dtszpowerlawphys & \dtszfourkpowerlawphys & \dtszfivekpowerlawphys& \dtszsixkpowerlawphys\\
free CIB+SZ& \dtsztwokmorecszphys & \dtszmorecszphys & \dtszfourkmorecszphys & \dtszfivekmorecszphys& \dtszsixkmorecszphys{} \\
\hline
 \rule{0pt}{4ex} & $D_{2000}^{\rm kSZ}$ &$D_{3000}^{\rm kSZ}$&$D_{4000}^{\rm kSZ}$&$D_{5000}^{\rm kSZ}$&$D_{6000}^{\rm kSZ}$ \\
\hline
Agora templates & - & \dkszagora &- &-&- \\ 
free CIB, $\ell^\alpha$ G15 SZ& \dksztwokpowerlawphys  & \dkszpowerlawphys{}& \dkszfourkpowerlawphys & \dkszfivekpowerlawphys& \dkszsixkpowerlawphys\\
free CIB+SZ & \dksztwokmorecszphys & \dkszmorecszphys & \dkszfourkmorecszphys & \dkszfivekmorecszphys& \dkszsixkmorecszphys\\
\hline
\end{tabular}
\caption{\label{tab:sz} 
The inferred tSZ power at 143\,GHz for angular multipoles $\ell \in [2000, 3000, ... 6000]$ and kSZ power at the same angular multipoles. 
Each row has the results for one model, labeled as in the legend of Fig.~\ref{fig:szspectra}. 
We highlight that the tabulated constraints can have correlated uncertainties.  
} \normalsize
\end{center}
\end{table*}

The kSZ power is best constrained between $\ell = 3500$ and $5500$, with uncertainties naturally increasing at low and high angular multipoles. 
We attribute the increase in uncertainty at $\ell\lesssim3500$ to confusion with the primary CMB.  
The uncertainties also grow at high $\ell$ for the same reasons that apply to the tSZ measurement: instrumental noise, high levels of CIB power and uncertainty in the tSZ-CIB correlation. 
Interestingly in the free CIB+SZ case, the inferred kSZ power rises to $\ell\sim 5000$ (though the statistical significance of this rise is low). 
We highlight  that the inferred kSZ power agrees at $1\,\sigma$ from $\ell=1700$ to 6000 between the two more-flexible modeling choices, while there are significant differences to the less-flexible template fit. 
While we hesitate to over-interpret the results at higher $\ell$ due to the relative magnitude of the CIB power, the recovered kSZ power falls towards zero at very small angular scales, and the upper limits at those small scales are consistent with either zero or the predictions from kSZ simulations.

\subsubsection{Dependence on CMB prior}\label{subsec:kszcmb}

A particular concern for the kSZ power is its potential degeneracy with the lensed, primary CMB anisotropy, as both signals have an identical frequency dependence. 
As noted in \S\ref{subsec:cmb}, we normally assume a Gaussian prior on the \lcdm{} parameters derived from the \textit{Planck} 2018 results \citep{planck18-6}. 
This Planck prior has been used for the kSZ and tSZ results just discussed.  
We now test what happens if priors based on the CMB-SPA and CMB-SPA-notau chains from \cite{camphuis25} are used instead.

\begin{figure}[t]\centering
\includegraphics[width=0.45\textwidth]{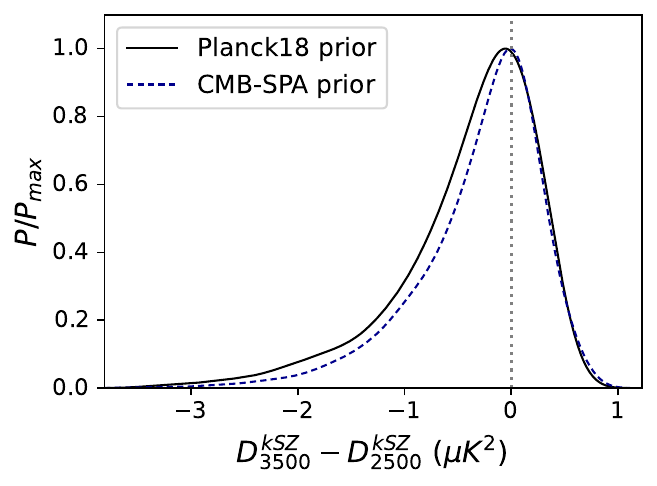}
\caption{  
1D  posteriors for the change in the total kSZ power between $\ell=3500$ and $\ell=2500$: $D_{3500}^{\rm kSZ} - D_{2500}^{\rm kSZ}$. 
The solid black line shows the baseline result for the free CIB, $\ell^\alpha$ G15 SZ model fit with priors on the \lcdm{} parameters based on the Planck 2018 results. 
The dashed blue line shows the effect of instead using priors on the  \lcdm{} parameters based the CMB-SPA chains. 
The two cases are consistent with each other. 
  }
  \label{fig:dksz1d}
\end{figure}

We see negligible shifts in the kSZ and tSZ power at $\ell=3000$ between the three \lcdm{} priors for both major model choices:   Agora templates and free CIB + power-law SZ.   
The median tSZ and kSZ powers at $\ell=3000$ shift by less than $0.2\,\sigma$ across the three CMB priors. 
The power-law exponents are also stable, with shifts of less that $0.1\,\sigma$ across the three CMB priors. 
We  show the inferred change in kSZ power between $\ell=2500$ and $\ell=3500$ for the baseline Planck 2018 prior and the CMB-SPA prior in Fig.~\ref{fig:dksz1d} (the CMB-SPA and CMB-SPA-notau results are very similar). 
The two posteriors are consistent with each other.

 \begin{figure*}[t]\centering
\includegraphics[width=0.45\textwidth]{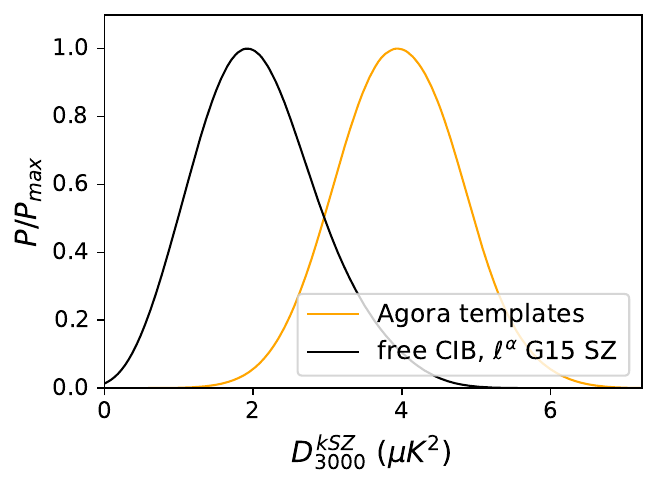}
\includegraphics[width=0.45\textwidth]{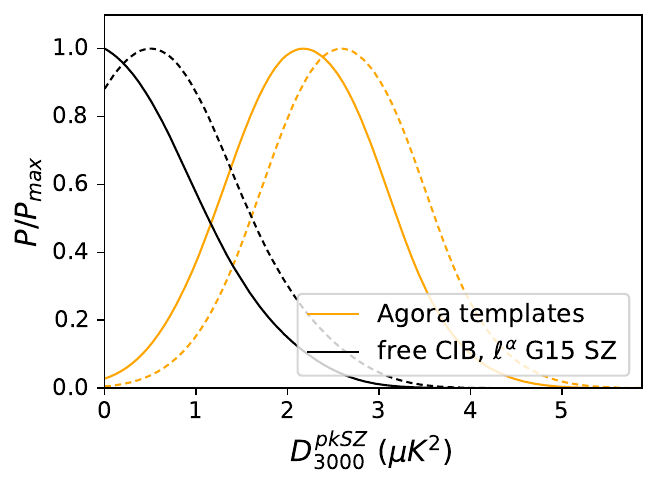}
\caption{  
\textbf{Left panel:} 1D  posteriors for the kSZ power at $\ell=3000$ (\dksz).
\textbf{Right panel:} 1D  posteriors for the implied patchy kSZ power at $\ell=3000$ (\dksz) after subtracting the predicted homogeneous kSZ power for $\sigma_8 = 0.812$ (solid lines) or $0.77$ (dashed lines). 
  }
  \label{fig:ksz1d}
\end{figure*}

 \begin{figure*}[t]\centering
\includegraphics[width=0.45\textwidth]{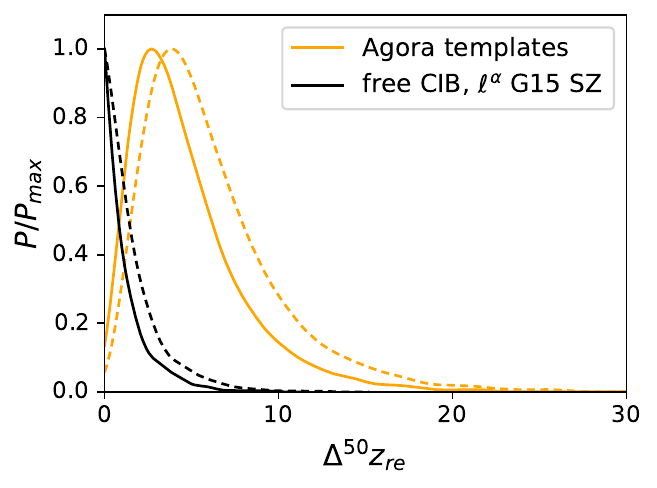}
\includegraphics[width=0.45\textwidth]{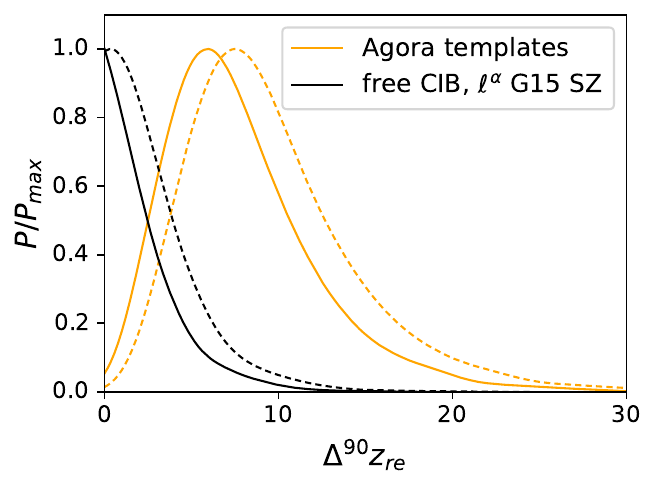}
\caption{  
1D  posteriors for the duration of reionization $\Delta z_{\rm re}$. 
The left panel uses the fitting formula for the patchy kSZ power at $\ell=3000$ reported by \cite{calabrese14}. 
Here the duration of reionization, $\Delta^{50} z_{\rm re}$, is  defined as the period between an ionization fraction of 0.25 to 0.75. 
The right panel uses the fitting formula from \cite{kramer25} based on the AMBER reionization simulation, in terms of a different period, $\Delta^{90} z_{\rm re}$, defined as the period  between ionization fractions of 0.05 to 0.95. 
In both panels, the solid lines are for the expected homogeneous kSZ power for the \textit{Planck} $\sigma_8= 0.812$ value, while the dashed lines show the effect if $\sigma_8=0.77$.
  }
  \label{fig:dz1d}
\end{figure*}

\subsubsection{Implications  for the epoch of reionization}

Measurements of kSZ power  provide information about the epoch of reionization. 
The observed kSZ power spectrum is the sum of patchy kSZ power from ionized bubbles during the epoch of reionization and  homogeneous kSZ power from the late-time universe. 
Thus to compare our measurement to the predictions of reionization models, we must split the measured kSZ power between the two sources. 
In this work, we  approach this splitting problem by simply subtracting a prediction for the homogeneous kSZ power at $\ell=3000$ of 
\be
\dhksz =1.65\left(\frac{\sigma_8}{0.8}\right)^{4.46}
\ee
from Eqn.~5 in \citenum{calabrese14} from the total kSZ power at that angular scale to estimate the patchy kSZ power. 
At the central value of $\sigma_8= 0.812$ for the Planck prior, the predicted homogeneous kSZ power is $\dhksz = 1.76\,\uksq$. 
We also show results for $\sigma_8 = 0.77$, which  leads to a lower estimate of $\dhksz = 1.35\,\uksq$. 
This second estimate serves two purposes. 
First it illustrates the dependence of the reionization results on the assumed homogeneous kSZ power, showing the effect of a $\sim$25\% change in the homogeneous  kSZ power. 
Second, the two choices bracket the current tension between measurements of $\sigma_8$ \citep{abdalla22}. 
For either homogeneous  kSZ prediction, we attribute the remaining kSZ power to the patchy kSZ power sourced during the epoch of reionization. 
In Fig.~\ref{fig:ksz1d}, we show the posteriors on the total kSZ power in the left panel, and the posteriors on the power attributed to the patchy kSZ effect in the right panel.

We then compare the patchy kSZ power to the predictions of two different reionization simulations, and derive constraints on reionization parameters in both cases. 
The first patchy kSZ prediction is taken from Eqn.~6 in  \citep{calabrese14}, who have used the models of  \citep{battaglia13a} to produce a fitting formula for the  patchy kSZ power at $\ell=3000$, \dpksz:
\begin{equation}
\dpksz = 2.03 \left[ \left(\frac{1+z_{\rm re}}{11}\right) - 0.12\right] \left(\frac{\Delta^{50} z_{\rm re}}{1.05}\right)^{0.51} \uksq.
\end{equation}
The redshift of reionization  $z_{\rm re}$ is defined as the redshift when the ionization fraction is 50\%, and the duration $\Delta^{50} z_{\rm re}$ as the period between 25\% and 75\% ionization fractions. 

The second estimate is taken from Table 4 of \citep{kramer25}, who derive power law scalings for the kSZ power in a bin $\ell\in [2950, 3050]$ using the semi-numerical AMBER simulations \citep{trac22}. 
The AMBER prediction for the patchy kSZ power is:
\begin{equation}
\dpksz = 1.75  \left(\frac{z_{\rm re}}{8.0} \right)^{1.4} \left(\frac{\Delta^{90} z_{\rm re}}{4.0}\right)^{0.75} \uksq.
\end{equation}
Here  $z_{\rm re}$ is defined as before while $\Delta^{90} z_{\rm re}$ is defined as the period between when the Universe was 5\% ionized to 95\% ionized. 
Importantly, the two predictions relate the kSZ power to different periods during reionization, either from 25\% to 75\% or from 5\% to 95\%, so the two sets of constraints are not directly comparable.

As discussed in \citetalias{reichardt21}, the uniform prior on kSZ power that is used in the MCMC sampling increases the posterior weight at small values of the duration of reionization. 
Thus we instead report results after resampling the chains to apply a uniform prior on either $\Delta^{50} z_{\rm re}$ or $\Delta^{90} z_{\rm re}$. 
As the allowed posterior range on optical depth is narrow, we also fix  $z_{\rm re}$ to the median value of 7.68 when post-processing chains to estimate the duration of reionization. 

We show the posteriors on the duration of reionization in Fig.~\ref{fig:dz1d}, with constraints on $\Delta^{50} z_{\rm re}$ based on the   fitting formula of \citep{calabrese14} in the left panel and constraints on $\Delta^{90} z_{\rm re}$ based on \citep{kramer25} in the right panel. 
For the estimated homogeneous kSZ power at the Planck $\sigma_8$ value, the data favor rapid epoch of reionization scenarios  with the more flexible models (solid black lines), and slower reionization with the Agora template fit (orange solid lines).
As expected, the estimated duration increases somewhat for the lower estimate of the homogeneous kSZ power at $\sigma_8=0.77$ (dashed lines). 
 
Looking at the posteriors for $\Delta^{50} z_{\rm re}$ in the left panel, we first note that the data favor a rapid ionization process from 25\% to 75\% ionization fraction in the free CIB + power-law SZ fit.
The data set a 95\% CL upper limit on the duration of $\Delta^{50} z_{\rm re}  \limdelzwhkszpriordzpowerlawphys$. 
The preferred duration increases in the Agora template fit, where the increased level of kSZ power leads to a median value of $\Delta^{50} z_{\rm re}=\delzwhkszpriordzagora$, or an upper limit of $\Delta^{50} z_{\rm re}  \limdelzwhkszpriordzagora$.
In both cases, the duration increases by $\sim$\,1 for the lower estimate of the homogenous kSZ power with $\sigma_8=0.77$. 

We compare these results to the recent limits on the duration of the reionization $\Delta^{50} z_{\rm re}$ from the kSZ trispectrum reported by \citep{raghunathan24}.  
In that work, the authors combine measurements of the kSZ trispectrum and  Gunn-Peterson trough to obtain an upper limit on the duration of reionization of $\Delta^{50} z_{\rm re}<4.5$ (95\% C.L). 
 As the homogeneous kSZ should contribute a larger fraction of the kSZ power than kSZ trispectrum,  the trispectrum results are expected to be less dependent on the homogeneous kSZ modeling. 
 The two kSZ analyses yield consistent constraints on the duration of reionization.

Finally, we look at the posteriors for the period from 5\% to 95\% ionization fraction shown in the right panel of Fig.~\ref{fig:dz1d}. 
These results depend  on the prediction for patchy kSZ power from \citep{kramer25} that is based on the AMBER simulations. 
In all cases, the implied 90\% durations are longer than the 50\% durations, as would be expected.  
For the Agora template fits, the inferred duration is $\Delta^{90} z_{\rm re}=\delzwhkszpriordzamberagora$. 
The median duration increases by 1.5 for the lower estimate of homogeneous kSZ power. 
The lower kSZ power in the more flexible power-law SZ template fits leads to an upper limit on $\Delta^{90} z_{\rm re}$ of 
 $\Delta^{90} z_{\rm re} \limdelzwhkszpriordzamberpowerlawphys$.

A caveat to the epoch of reionization constraints reported here is that the constraints only use the measured kSZ power at $\ell= 3000$, and do not use  kSZ information from other angular scales. 
It would be interesting to fit the kSZ power at all angular scales to a full model for the homogeneous and patchy kSZ power spectra to derive reionization constraints that incorporate all information. 
However, such an analysis is beyond the scope of this work.

\subsection{Constraints on the tSZ-CIB correlation}\label{subsec:szcibparams}

Both the tSZ and CIB are tracers of large scale structure, and are thus expected to be spatially correlated at some level. 
The resulting tSZ-CIB cross-spectra have a similar spectral dependence, but opposite sign, to the kSZ when considering the 150$\times$150\,GHz, 95$\times$150\,GHz, and 150$\times$220\,GHz cross-spectra, while differing more significantly from the kSZ in the other frequency pairs. 
As a result,  an increase in the tSZ-CIB correlation leads to an increase in inferred kSZ power. 

With the Agora template fits, we find a preference for a positive spatial correlation at $\ell=3000$ of $\xi_{\rm 3000} =\tszcibagora$. 
The correlation drops for the free CIB + power-law G15 SZ model to $\xi_{\rm 3000} = \tszcibpowerlawphys$  while the free CIB+SZ model splits the difference at   $\xi_{\rm 3000} = \tszcibmorecszphys$.
At this angular scale, the recovered correlation  is consistent at $2.2\,\sigma$ between the three main model options. 

\begin{figure}[t]\centering
\includegraphics[width=0.48\textwidth]{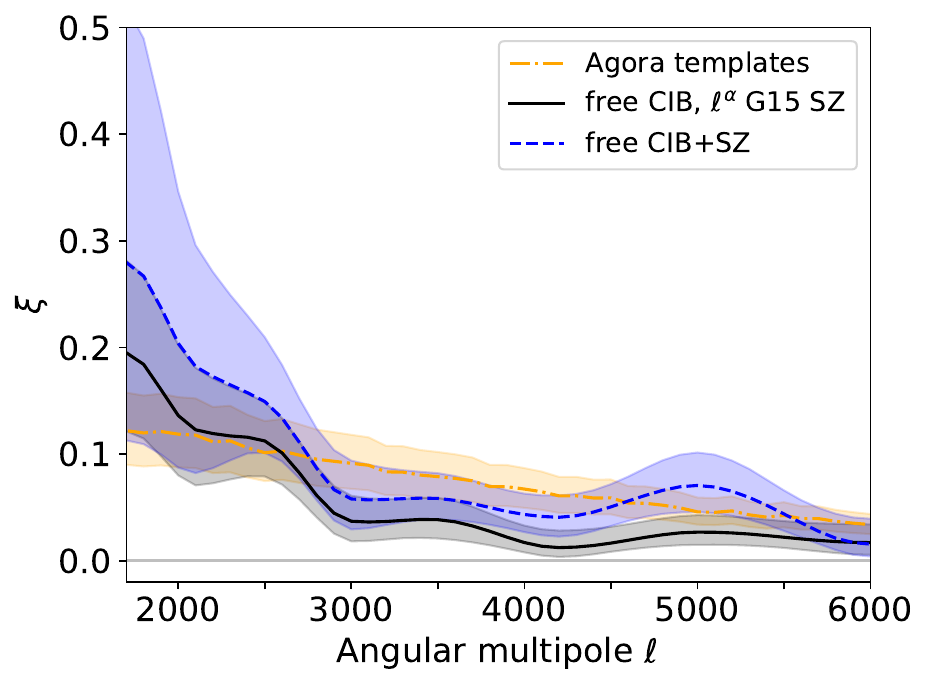}
\caption{   
The median values and 68\% confidence regions for the tSZ-CIB correlation $\xi$ across the same set of modeling options as Fig. \ref{fig:szspectra}.
Results from the Agora kSZ, tSZ, and tSZ-CIB correlation template fits are marked by the orange dashed line. 
The result with the free CIB model and the SZ terms described by G15 SZ templates scaled by free $\ell^\alpha$ terms is shown by the black line. 
The free CIB+SZ case is shown by the blue dashed line. 
The three cases agree at $\sim2\sigma$ across these angular scales. 
  }
  \label{fig:tszcibpower}
\end{figure}

The angular dependence of the recovered tSZ-CIB correlation is plotted in Fig.~\ref{fig:tszcibpower}.
The orange dash-dot line shows the single-parameter fit with the Agora templates, while the two free CIB models (with eight parameters for the tSZ-CIB correlation) are shown by the black and blue lines. 
The recovered tSZ-CIB correlation has some dependence on the SZ modeling, but there is qualitative agreement between the cases. 
The data favor a tSZ-CIB correlation that is broadly decreasing with increasing $\ell$. 

This fall-off with $\ell$ of tSZ-CIB correlation is the opposite of the rising $\ell$-dependence predicted by the Z12 tSZ-CIB template used by G15. 
This mismatch drives the inferred correlation towards zero for the G15 template fits. 
The effectively zero tSZ-CIB correlation with the G15 shapes also explains much of the shift in the recovered kSZ and tSZ power between G15 and Agora templates that is discussed in \S\ref{subsec:sz}.

\subsection{Cosmic infrared background}

\begin{figure*}[t]\centering
\includegraphics[width=0.45\textwidth]{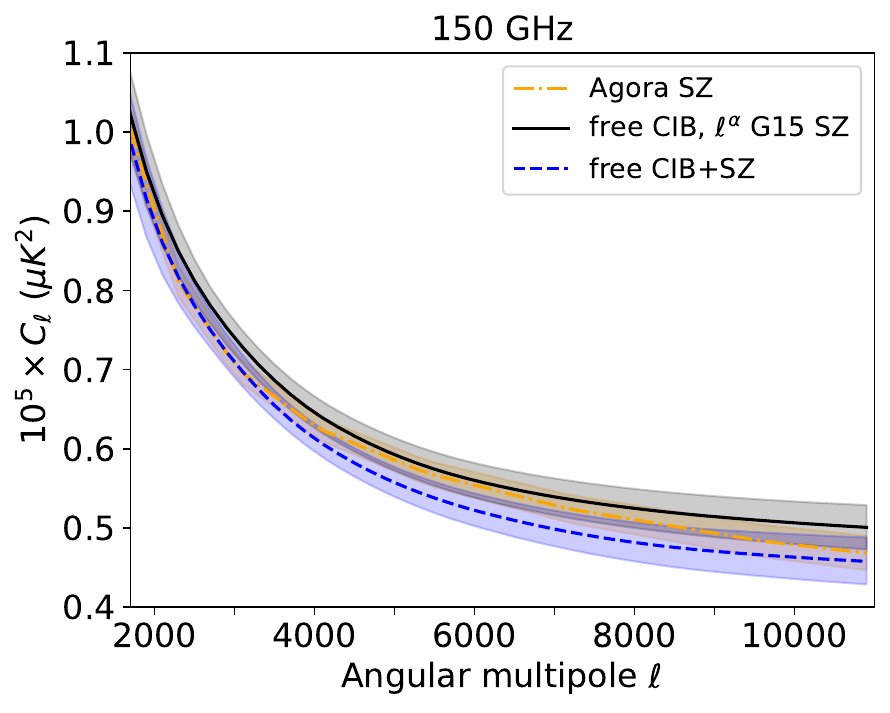}
\includegraphics[width=0.45\textwidth]{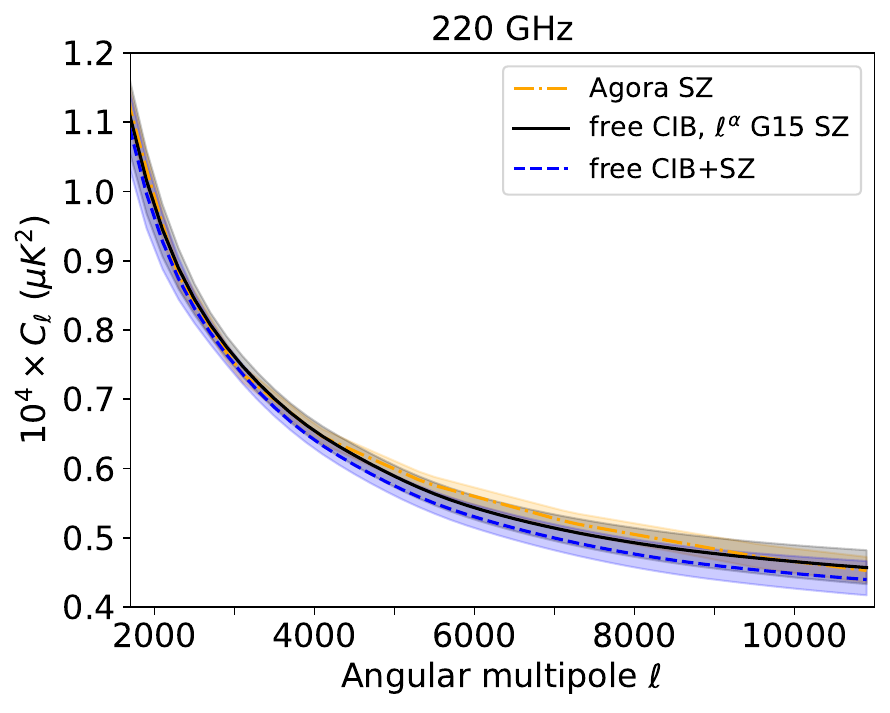}
\caption{   
CIB power at 150 GHz (left) and 220\,GHz (right) for three model choices. 
The median power is marked by each line with shading showing the 68\% confidence region. 
The CIB power is robust to the modeling options at 150 and 220\,GHz (where it accounts for most of the observed power), with the largest variations at $6000\le \ell \le 9000$. 
  }
  \label{fig:cibpower}
\end{figure*}

The cosmic infrared background is the largest signal after the primary CMB at 150 and 220 GHz, and has a small but non-negligible contribution at 95\,GHz. 
Naturally,  the CIB model is also critically important to fitting the bandpowers,  as seen  in Tab.~\ref{tab:chisq} where dropping the CIB terms generally leads to increases in the best-fit \chisq{} by $\sim$200-400 per dof. 
Given this, 
while we see little evidence the 1- and 2-halo template fits used by R21 are a poor description of the clustered CIB power, we explore the results for both the 1- and 2-halo template approach and the non-template `free CIB' approach. 
We focus on three model cases: the Agora template fits, the free CIB + power-law G15 SZ case, and the free CIB+SZ case.

We find the recovered CIB power in the models is fairly stable. 
Fractionally, the CIB power is most tightly constrained at 220 GHz, and most weakly at 95\,GHz, as one would expect given the steeply rising nature of the modified blackbody SED. 
The total CIB power at $\ell=3000$ is approximately 10\,\uksq{} at 150\,GHz and 108\,\uksq{} at 220\,GHz. 
Specifically at $\ell=3000$ and 150\,GHz, the total CIB power in the three considered cases is $D_{3000}^{\rm CIB} = \dcibtotonefiftyagora{}$, \dcibtotonefiftypowerlawphys{}, and \dcibtotonefiftymorecszphys{} for the Agora templates, free CIB + power-law SZ, and CIB+SZ-free cases respectively. 
At 220\,GHz, the total CIB power in the three considered cases is $D_{3000}^{\rm CIB} = \dcibtottwotwentyagora{}$, \dcibtottwotwentypowerlawphys{}, and \dcibtottwotwentymorecszphys{} for the same three cases. 
At $\ell=3000$, the inferred CIB power levels differ by less than  1\,$\sigma$ across the model choices for both observing frequencies. 
The total 150 and 220\,GHz CIB power at all multipoles for different cases is plotted in the left and right panels respectively of Fig.~\ref{fig:cibpower}.

While the total power is stable,  there are statistically significant differences in the individual model parameters between the Agora template and free CIB approaches. 
The  \dcibtotonefiftyagora{} of CIB power at 150\,GHz noted above in the Agora template fits is primarily in the 1-halo term. 
The Poisson power level is $\dcibponefiftyagora$, while the 1- and 2-halo power levels are \dcibclaonefiftyagora{} and \dcibclbonefiftyagora{} respectively. 
The two free CIB cases increase the Poisson power while decreasing the total clustered power at $\ell=3000$. 
For the free CIB + power-law SZ fit, the Poisson power at 150\,GHz is $D_{3000}^{\rm CIB~pois} =  \dcibponefiftypowerlawphys{}$ while the clustered power is $D_{3000}^{\rm CIB~cl} = \dcibclonefiftypowerlawphys{}$. 
The inferred Poisson and clustered CIB power between the two free CIB cases agrees to $1\,\sigma$.
 The free CIB fits at $\ell=3000$ and 150\,GHz are robust to the SZ model chosen.

The picture is similar at 220\,GHz. 
Again, while the total power is quite stable, the Agora template fits favor less Poisson power and more clustered power than the free CIB fits. 
For the Agora fits at 220\,GHz, the Poisson power level is $\dcibptwotwentyagora$, while the 1- and 2-halo power levels are \dcibclatwotwentyagora{} and \dcibclbtwotwentyagora{} respectively. 
The free CIB fits prefer more Poisson power, finding $D_{3000}^{\rm CIB~pois} = \dcibptwotwentypowerlawphys$ for the Poisson power and  $D_{3000}^{\rm CIB~cl} =\dcibcltwotwentypowerlawphys$ for the clustered power.
 These numbers are for the free CIB+$\ell^\alpha$ SZ case; the results for the CIB+SZ-free case agree at $< 0.4\,\sigma$.
As at 150\,GHz, the free CIB model posteriors are robust to the SZ modeling. 
We highlight that while the allocation between clustered and Poisson power changes between the model choices, the implied total CIB powers at both frequencies are  robust and agree within $2\,\sigma$ across the entire multipole range. 

The implied CIB power at 95\,GHz and $\ell=3000$ is $\dcibtotninetyfivepowerlawphys$ for the free CIB+$\ell^\alpha$ SZ case, with the other two cases agreeing at $\lesssim 0.2\,\sigma$. 
We interpret this stability as a consequence of the simple SED with one free parameter $\beta$ that can be completely determined by measurements at 150 and 220\,GHz rather than a robust independent detection of CIB power at 95\,GHz.  
We expect the 95\,GHz model power would show more variation with a more complex galaxy SED model. 
The CIB level at 95\,GHz  is smaller than all other model components at $\ell=3000$, including the kSZ effect.

\subsection{Radio galaxy power}

The data well constrain the power due to a Poisson distribution of radio galaxies at 95 and 150\,GHz.
The inferred level of radio power is qualitatively similar between the different model choices, with agreement at the 10\% (15\%) level at 95\,GHz (150\,GHz) across the set of models. 
The model choices that lead to the largest changes in the inferred radio power are the tSZ spectrum shape and the CIB and tSZ-CIB modeling.

At 95\,GHz, we recover $D_{3000}^{\rm rg, ~95\,GHz} = \drgninetyfiveagora$, \drgninetyfivepowerlawphys{}, and  \drgninetyfivemorecszphys{} for the Agora templates, free CIB + $\ell^\alpha$ SZ, and free CIB+SZ cases respectively. 
As expected, the radio power is lower in the 150\,GHz autospectrum where the posterior values are $D_{3000}^{\rm rg, ~150\,GHz} = \drgonefiftyagora$, \drgonefiftypowerlawphys{}, and \drgonefiftymorecszphys{} for the same three cases. 
For the flux masking threshold of 6 mJy in this work, the source count model from  \citep{dezotti10}  predicts $D_{3000}^{\rm rg, ~150\,GHz} = 1.2\,\mu {\rm K}^2$, which is consistent with the data's preference.

There is a signficant degeneracy between the average spectral index $\alpha_{\rm rg}$ and variance $\sigma^2_{\rm rg}$. 
This is expected as one can produce a similar power ratio between 95 and 150\,GHz by simultaneously increasing $\alpha_{\rm rg}$ and $\sigma^2_{\rm rg}$.


\section{Conclusion}
\label{sec:conclusion}

We have presented bandpowers measured from 1646\,\sqdeg{} of the SPT-3G survey maps at 95, 150, and 220\,GHz, using data collected during the 2019 and 2020 austral winters. 
Compared to previous South Pole Telescope measurements, these data reduce uncertainties on the millimeter sky power at these frequencies and angular scales by up to a factor of 8, and they represent the best current measurement in all three observing bands at $\ell \gtrsim 3000$.

 We explore several different options to model the observed power to understand the model dependence of the recovered posteriors. 
 We find the template fits sufficient for previous measurements with SPTpol and SPT-SZ yield kSZ and tSZ results that are highly template-dependent at the SPT-3G sensitivity. 
 Varying the templates assumed for the tSZ-CIB correlation and to a lesser extent the kSZ and tSZ power spectra can change the median values of the kSZ and tSZ power, as well as the tSZ-CIB correlation, by up to $5\,\sigma$. 
 The PTE of the best-fit point for the template fits is also extremely poor, indicating these models are inadequate to fully explain the bandpowers presented in this work. 
This new sensitivity  means the  data can internally constrain the angular dependence of these terms without depending on simulation-based templates. 
 In the free SZ case, we present $>2\,\sigma$ measurements of the tSZ power in five bins over the range $2000 < \ell < 6000$ and of the kSZ power in four bins over the range $3000 < \ell < 6000$.

 We find the tSZ power levels inferred at $2500<\ell<3500$ to be stable across the models considered, with more model dependence on the recovered kSZ power. 
 These are also the angular scales with the smallest uncertainties on the tSZ power, consistent with the decisions in previous works to evaluate the SZ power levels at $\ell=3000$. 
 When adding two degrees of freedom to the SZ modeling through prefactors of the form $\ell^{\alpha_{\rm kSZ}}$ and  $\ell^{\alpha_{\rm tSZ}}$ respectively, we recover  $\dtsz = \dtszpowerlawphys$ and $\dksz =\dkszpowerlawphys$ for the power at $\ell=3000$.  
 However as seen in Fig.~\ref{fig:szspectra},  the differences between models in the recovered tSZ power  grow towards smaller angular scales.

Under strong assumptions about the allocation of the kSZ power between the homogeneous and patchy kSZ terms,  we use the measured kSZ power at $\ell=3000$ to derive constraints on the duration of the epoch of reionization. 
As we detect low-levels of positive patchy kSZ power for both homogeneous kSZ estimates considered, the data favor short, but non-zero, durations of reionization. 
For our fiducial estimate of the homogeneous kSZ power and the $\ell^\alpha$ SZ modeling, we find a duration of reionization from 25\% to 75\% ionization fraction  of  $\Delta^{50} z_{\rm re}=\delzwhkszpriordzpowerlawphys$ when using the fitting formula of \citep{calabrese14}  for the patchy kSZ power, which are based on the  simulations in \citep{battaglia13a}. 
Instead, using the fitting formula of \citep{kramer25} which are based on the AMBER simulations \citep{trac22} and use a different definition of the duration of reionization, we find a duration from 5\% to 95\% ionization fraction of $\Delta^{90} z_{\rm re}=\delzwhkszpriordzamberpowerlawphys$.

We find evidence for a positive spatial correlation between the sources contributing to the CIB and tSZ power spectra, as is expected from simulations and previous measurements. 
The data favor a correlation which is generally falling from $\ell=1700$ to 6000 and only set an upper limit on the correlation at higher $\ell$. 
However, the exact levels of correlation vary between the SZ model options considered with lower correlation at a given angular scale increasing the kSZ power while reducing the tSZ power at that scale.

These bandpowers are the first measurements at these angular scales from the SPT-3G survey. 
The SPT-3G survey will continue to gather data through at least 2027, and the full depth SPT-3G survey maps of this survey field are expected to reduce the bandpower uncertainties from this work by a factor of 3.5 in the noise-dominated regime. 
The full SPT-3G survey also covers an additional 8400\,\sqdeg{} of sky, which in combination should reduce uncertainties by a factor of 2.5 in the signal-dominated regime. 
Even tighter constraints on the CMB secondary anisotropies and CIB will come from the next generation of CMB experiments such as SPT-3G+, SO, CMB-S4, or CMB-HD \citep{anderson22,simonsobservatorycollab19,cmbs4collab19,sehgal19}. 
 Interpreting the precise measurements of the mm-wave sky from these experiments will require advances in our modelling of the extragalactic foreground components and their correlations. 

\acknowledgments
The South Pole Telescope program is supported by the National Science Foundation (NSF) through awards OPP-1852617 and OPP-2332483. Partial support is also provided by the Kavli Institute of Cosmological Physics at the University of Chicago. 
Argonne National Laboratory’s work was supported by the U.S. Department of Energy, Office of High Energy Physics, under contract DE-AC02-06CH11357. 
The UC Davis group acknowledges support from Michael and Ester Vaida. 
Work at the Fermi National Accelerator Laboratory (Fermilab), a U.S. Department of Energy, Office of Science, Office of High Energy Physics HEP User Facility, is managed by Fermi Forward Discovery Group, LLC, acting under Contract No. 89243024CSC000002.
The Melbourne authors acknowledge support from the Australian Research Council’s Discovery Project scheme (No. DP210102386). 
The Paris group has received funding from the European Research Council (ERC) under the European Union’s Horizon 2020 research and innovation program (grant agreement No 101001897), and funding from the Centre National d’Etudes Spatiales. 
The SLAC group is supported in part by the Department of Energy at SLAC National Accelerator Laboratory, under contract DE-AC02-76SF00515.

Some of the results in this paper have been derived using the HEALPix \citep{gorski05} package.

\bibliographystyle{JHEP}
\bibliography{../../BIBTEX/spt_1995_to_2000.bib,../../BIBTEX/spt_2000_to_2005.bib,../../BIBTEX/spt_2005_to_2010.bib,../../BIBTEX/spt_2010_to_2015.bib,../../BIBTEX/spt_2015_to_2020.bib,../../BIBTEX/spt_2020_to_2025.bib,../../BIBTEX/spt_2025_and_after.bib,../../BIBTEX/spt_before_1995.bib,unpublished.bib}

\end{document}